\documentclass[aps, pra, reprint, superscriptaddress]{revtex4-2}
\usepackage{graphicx} 
\usepackage{amsmath, physics}
\usepackage{amssymb}
\usepackage{amsfonts}
\usepackage[unicode]{hyperref}
\hypersetup{
   unicode=true,          %
   plainpages=false,
   colorlinks=true,       
   citecolor=blue,        
   urlcolor=blue
}
\usepackage[caption=false,position=top,singlelinecheck=off,justification=raggedright]{subfig}
\usepackage{color}

\begin{document} 
\title{Controlled bit flip of period-doubling and discrete time crystalline states in open systems}

\author{Roy D. Jara Jr.}
\email{rjara@nip.upd.edu.ph}
\affiliation{National Institute of Physics, University of the Philippines, Diliman, Quezon City 1101, Philippines}

\author{Phatthamon Kongkhambut}
\affiliation{Center for Optical Quantum Technologies and Institute for Quantum Physics, Universit\"{a}t Hamburg, 22761 Hamburg, Germany}
\affiliation{Quantum Simulation Research Laboratory, Department of Physics and Materials Science, Faculty of Science, Chiang Mai University, Chiang Mai, 50200, Thailand}
\affiliation{Thailand Center of Excellence in Physics, Office of the Permanent Secretary, Ministry of Higher Education, Science, Research and Innovation, Thailand}

\author{Hans Ke{\ss}ler}
\affiliation{Center for Optical Quantum Technologies and Institute for Quantum Physics, Universit\"{a}t Hamburg, 22761 Hamburg, Germany}

\author{Andreas Hemmerich}
\affiliation{Center for Optical Quantum Technologies and Institute for Quantum Physics, Universit\"{a}t Hamburg, 22761 Hamburg, Germany}
\affiliation{The Hamburg Center for Ultrafast Imaging, Luruper Chaussee 149, 22761 Hamburg, Germany}

\author{Jayson G. Cosme}
\email{jcosme@nip.upd.edu.ph}
\affiliation{National Institute of Physics, University of the Philippines, Diliman, Quezon City 1101, Philippines}


\begin{abstract}

In this work, we explore the robustness of a bit-flip operation against thermal and quantum noise for bits represented by the symmetry-broken pairs of the period-doubled states in a classical parametric oscillator and discrete time crystal (DTC) states in an open fully-connected spin-cavity system, respectively. The bit-flip operation corresponds to switching between the two period-doubled and DTC states induced by a defect in a periodic drive. The defect is introduced in a controlled manner by linearly ramping the phase of the modulation of the drive. In the absence of stochastic noise, strong dissipation results in a more robust bit-flip operation in which slight changes to the defect parameters do not significantly lower the success rate of bit-flips. The operation remains robust even in the presence of stochastic noise when the defect duration is sufficiently large.  By considering parameter regimes in which the DTC states in the spin-cavity system do not directly map to the period-doubled states, we reveal that this robustness stems from the system being quenched by the defect towards a new phase that has enough excitation to suppress the effects of the stochastic noise. This allows for precise control of the bit-flip operations by tuning into the preferred intermediate state that the system will enter during a bit-flip operation. We demonstrate this in a modified protocol based on precise quenches of the driving frequency.

\end{abstract}

\maketitle

\section{Introduction}

The simplest system that can break discrete time translation symmetry is a single parametric oscillator (PO). When driven resonantly, it enters a period-doubled state characterized by a subharmonic oscillation relative to the drive accompanied by an exponential growth of the system's response amplitude \cite{kovacic_mathieus_2018}. In the case of a nonlinear oscillator, this exponential growth is tapered by the nonlinearity, forcing the response amplitude to relax to a constant value \cite{kovacic_mathieus_2018}. Similar to systems that spontaneously break discrete symmetries in equilibrium, when a PO enters a period-doubled state, the system randomly chooses one of its two degenerate states, distinguished by a shift of $\pi$ on their oscillation phase \cite{kovacic_mathieus_2018, leuch_parametric_2016}. Due to its simplicity, the PO can be emulated in a wide range of setups \cite{bello_persistent_2019, mahboob_bit_2008, fabiani_parametrically_2022, elyasi_stochasticity_2022, leuch_parametric_2016, nosan_gate_controlled_2019, apffel_experimental_2024, frimmer_rapid_2019}, with the driven pendulum being the simplest example \cite{kovacic_mathieus_2018}.

\begin{figure}
\centering
\includegraphics[scale=0.375]{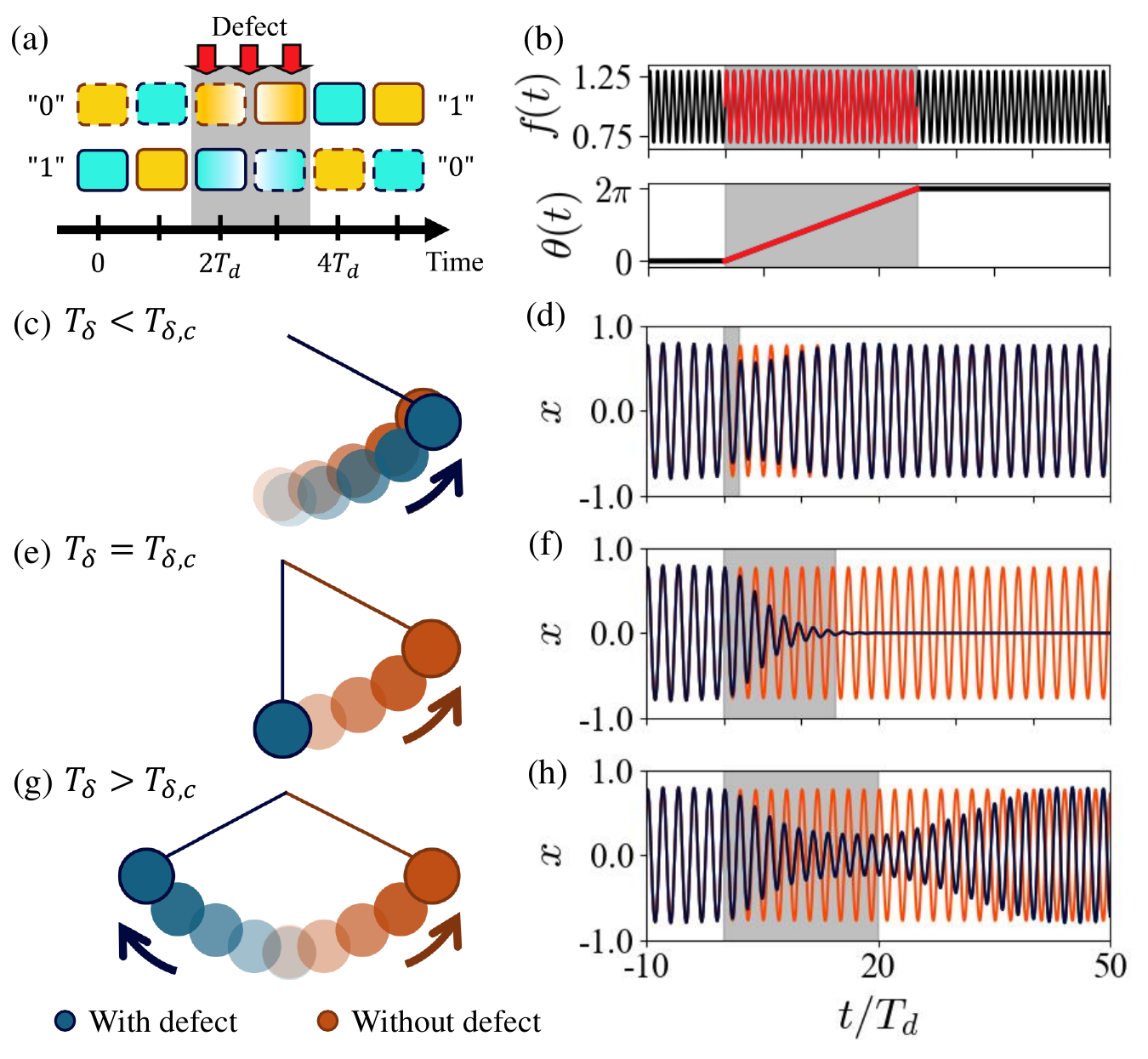}
\caption{(a) Sketch of a bit-flip operation. Two degenerate dynamical states, labeled as ``0" and ``1", are flipped onto their symmetry-broken pairs by applying a defect protocol within some time duration $T_{\delta}$. (b) (Top panel) Periodic drive with a defect protocol applied during the time window $t=0$ and $t = T_{\delta} = 25 T_{d}$. (Bottom panel) Phase ramp protocol corresponding to the defect applied in the top panel. (c)-(h) Sketch and exemplary dynamics of a PO after a defect protocol for (c), (d) $T_{\delta} < T_{\delta, c}$, (e), (f) $T_{\delta} = T_{\delta, c}$, and (g), (h) $T_{\delta} > T_{\delta, c}$. The gray regions correspond to the duration at which the defect protocol is applied. The rest of the parameters are $\omega_{d}=2 \Omega$, $A = 0.3\Omega$, and $\gamma = 0.1 \Omega$.}
\label{fig:schematics}
\end{figure}

Due to the degenerate nature of the period-doubled states, POs have been considered as a good candidate for emulating classical bits \cite{goto_parametron_1959, heugel_ising_2022, calvanese_strinati_theory_2019, ameye_parametric_2025, mahboob_bit_2008, fabiani_parametrically_2022, elyasi_stochasticity_2022, leuch_parametric_2016, nosan_gate_controlled_2019}, motivating the search for robust methods to manipulate these states. It has been shown in Refs.~\cite{leuch_parametric_2016, frimmer_rapid_2019, nosan_gate_controlled_2019} that POs can switch states by varying the natural frequency of the oscillator within a time duration, demonstrating the possibility of performing the simplest classical bit-operation on this system: a bit-flip. For systems with inaccessible natural frequency, however, Ref.~\cite{apffel_experimental_2024} has introduced a method for controlled switching between period-doubled states by applying a defect on the system's drive as shown in Fig.~\ref{fig:schematics}(b). The defect protocol consists of linearly ramping the phase of the drive, $\theta(t)$, from $\theta = 0$ to $2\pi$ within a duration of $T_{\delta}$. Depending on the chosen $T_{\delta}$, the defect protocol can be effectively used as a bit-flip operation for bits encoded in the period-doubled states, as shown in Figs.~\ref{fig:schematics}(c)-\ref{fig:schematics}(g). While Ref.~\cite{apffel_experimental_2024} has demonstrated the robustness of this method in noiseless oscillators with weak dissipation, it remains an open question whether it will persist for arbitrary dissipation strength and in the presence of random fluctuations.

Periodically driven many-body systems can also spontaneously break discrete time translation symmetry \cite{else_discrete_2020}. In this case, their collective behavior results in the emergence of a discrete time crystal (DTC). It manifests as a subharmonic oscillation of an order parameter, with an oscillation phase chosen from the multiple degenerate states associated with the broken symmetry \cite{else_discrete_2020, smits_spontaneous_2021, munoz-arias_floquet_2022}. In the case of DTCs with period-doubling oscillations, the system picks an oscillation phase from the two degenerate states owing from the broken $\mathbb{Z}_{2}$ symmetry \cite{smits_spontaneous_2021, munoz-arias_floquet_2022}, akin to the period-doubled states of POs. The DTCs have been extensively studied under wide range of platforms, ranging from, but not limited to, networks of classical oscillators \cite{heugel_classical_2019, yao_classical_2020, nicolaou_anharmonic_2021, heugel_role_2023, yi-thomas_theory_2024}, spin systems \cite{FloquetLMG, switzer_realization_2025, euler_metronome_2024, Lazarides_2020, II_von, Khemani_PRL, frey_realization_2022, FLoquet, pizzi_higher-order_2021, nurwantoro_discrete_2019, munoz-arias_floquet_2022}, Rydberg atoms \cite{liu_higher-order_2024, fan_discrete_2020}, bosonic systems \cite{autti_ac_2021, bakker_driven-dissipative_2022, yang_dynamical_2021, chitra_dynamical_2015, zhu_dicke_2019, gong_discrete_2018, nie_mode_2023, jager_dissipative_2023, kongkhambut_realization_2021, skulte_parametrically_2021, kesler_observation_2021, tuquero_dissipative_2022, cosme_time_2019}, superconductors \cite{Ojeda_2023, ojeda_collado_emergent_2021, homann_higgs_2020}, and particles under oscillating mediums \cite{kuros_phase_2020, giergiel_creating_2020, simula_droplet_2023}.

In Ref.~\cite{jara_jr_theory_2024}, it has been shown that there is a correspondence between the degenerate states of the DTCs, which we will refer to as DTC states, in periodically driven fully-connected spin-cavity systems and the period-doubled states of coupled POs. This mapping opens the possibility of manipulating DTC states using the methods for period-doubled states. Indeed, this has been explored for DTC states in other systems, such as closed, integrable bosonic system \cite{yang_dynamical_2021} and classical time crystals \cite{simula_topological_2024, dhardemare_probing_2020}. Due to the non-integrability of fully-connected spin-cavity systems, however, such systems generally heat up, leading to a finite lifetime for the DTCs. While this problem can be circumvented by connecting the system to a bath that will serve as a heat sink \cite{cosme_bridging_2023}, it renders the spin-cavity system susceptible to stochastic noise from the environment \cite{mivehvar_cavity_2021, tuquero_impact_2024, cenedese_thermo_2025}. This poses the question of whether we can perform bit-flip operations on the DTC states of fully-connected open spin-cavity systems, and how robust the operation would be against quantum fluctuations.

In this work, we address the robustness of bit-flip operations on the period-doubled states of POs connected to a thermal bath. We then apply these results to demonstrate that robust-bit-flip operations are possible for DTC states of open spin-cavity systems both in the thermodynamic limit and finite-size limit, where quantum fluctuations become dominant. We also consider system parameters leading to DTC states that do not have any direct mapping to period-doubled states. In this case, we demonstrate that the defect protocol can be viewed as a sudden quench towards new dynamical phases. This opens the possibility for more precise control of bit-flip operations on the DTC states.

The paper is structured as follows. In Sec.~\ref{sec:system_of_interest}, we introduce the systems of interest together with the protocol for the bit-flip. We then present in Sec.~\ref{sec:parametric_state_swithing} the results for the robustness of the bit-flip operations for arbitrary dissipation without noise, and fixed dissipation with arbitrary noise strength. We also demonstrate in this section the applicability of the bit-flip operation on DTC states with no direct mapping to period-doubled states, and give insights on the dynamics during the defect and their implications on the protocol's robustness against noise. In Sec.~\ref{sec:modified_defects}, we consider other variations of the defect protocol beyond the phase ramp protocol shown in Fig.~\ref{fig:schematics}(b) and demonstrate their robustness in performing bit-flips given a set of defect parameters. Finally, we provide a summary and conclusion in Sec.~\ref{sec:conclusion}.

\section{Systems of interest}\label{sec:system_of_interest}

\begin{figure*}[!htbp]
\centering
\includegraphics[scale=0.5]{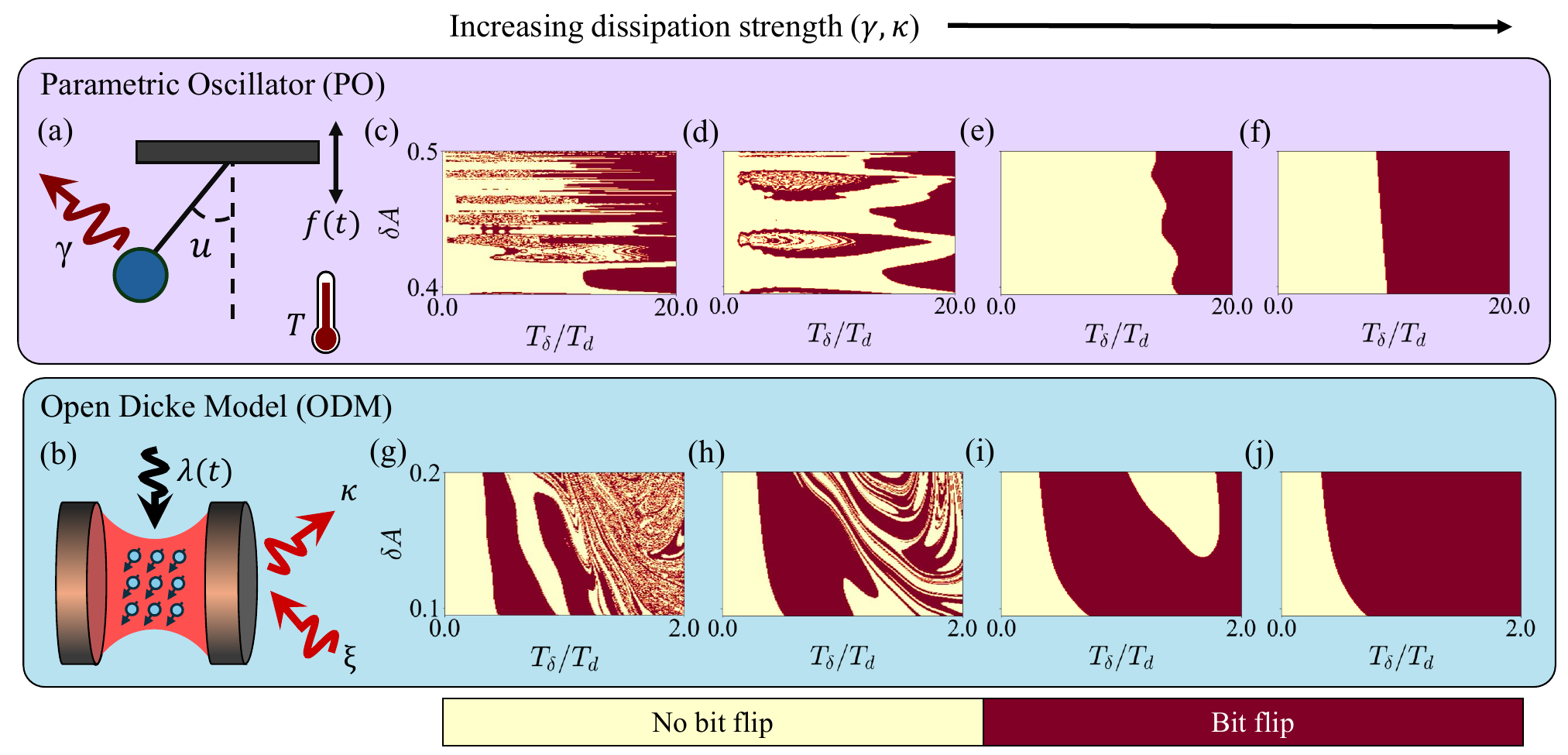}
\caption{Sketch of (a) a parametric oscillator connected to a thermal bath and (b) a finite-sized open Dicke model. (c)--(f) Zero-temperature bit-flip diagram of the PO driven at resonance as a function of the defect duration $T_{\delta}$ and amplitude detuning $\delta A = A - A_{r}$ for increasing dissipation strength, (c) $\gamma = 0.005 \Omega$, (d) $\gamma = 0.01\Omega$, (e) $\gamma = 0.025 \Omega$, and (f) $\gamma = 0.1\Omega$. (g)-(j) Bit-flip diagram of the resonantly driven ODM in the thermodynamic limit for (g) $\kappa = 0.01\omega$, (h) $\kappa = 0.02 \omega$, (i) $\kappa = 0.06 \omega$, (j) $\kappa = 0.1\omega$. The remaining parameters for the ODM are $\lambda_{0} = 0.9\lambda_{c}$, and $\omega_{0} = \omega$.}
\label{fig:switch_diagram}
\end{figure*}

As a test bed for exploring the robustness of bit-flips on period-doubled states, we consider a PO of mass $m$ and length $L$ coupled to a thermal bath, depicted in Fig.~\ref{fig:switch_diagram}(a). The system is described by the Langevin equation \cite{Kampen1992}
\begin{equation}
\label{eq:stochastic_pendulum}
\frac{d^{2} u}{d t^{2}}  + \gamma \frac{d u}{d t} + \Omega^{2} f(t) \sin u = \eta(t),
\end{equation}
where $\Omega = \sqrt{L/g}$ is the natural frequency of the pendulum, $g$ is the acceleration due to the gravity, and $u$ is its angular position. The system is driven by  
\begin{equation}
f(t) = 1 + A\sin \left(\omega_{d}t  + \theta\right),
\end{equation}
with a frequency $\omega_{d}$, an amplitude $A$, and a phase $\theta$. The dissipation strength of the pendulum is given by $\gamma$, while $\eta(t)$ corresponds to the thermal noise due to a bath. This stochastic noise satisfies the following conditions \cite{Kampen1992},
\begin{equation}
\label{eq:po_noise_condition}
\left< \eta(t) \right> = 0, \quad \left< \eta(t)\eta(t') \right> = 2 \tilde{T} \Omega^{2} \gamma \delta\left( t - t^{'} \right),
\end{equation}
where $\tilde{T}$ is the dimensionless temperature defined as
\begin{equation}
\tilde{T} = \frac{k_{B}T}{mL^{2}\Omega^{2}},
\end{equation} 
with $k_{B}$ being the Boltzmann constant.

In the zero-temperature limit, $\tilde{T} \rightarrow 0$, the PO can be driven resonantly by setting $\omega_{d}$ and $A$ to \cite{kovacic_mathieus_2018, heugel_ising_2022} (see Appendix \ref{subsec:resonance_condition} for more details),
\begin{equation}
\label{eq:dpp_resonant_condition}
\omega_{r, \mathrm{PO}} = 2\Omega, \quad A_{r, \mathrm{PO}} = 2\gamma / \Omega,
\end{equation}
where $\omega_{r, \mathrm{PO}}$ is the primary resonant frequency of the PO and $A_{r, \mathrm{PO}}$ is the minimum $A$ needed to enter a period-doubled state at $\omega_{r, \mathrm{PO}}$.  In this case, the system exhibits a period-doubled oscillation with respect to the driving period, $T_{d}$, with an oscillation phase spontaneously chosen from the two degenerate period-doubled states \cite{kovacic_mathieus_2018, leuch_parametric_2016}. This phase shift, which we refer to as the absolute-time phase $\varphi$, can be extracted from an order parameter, $O(t)$, capturing the system's period-doubling response using the equation \cite{simula_topological_2024} 
\begin{equation}
\label{eq:complex_amplitude}
O_{R}(t) = R(t)e^{i\varphi(t)} = \frac{\omega_{R}}{\pi} \int_{t}^{t + \frac{2\pi}{\omega_{R}}} e^{i\omega_{R}\tau}O(\tau)d\tau,
\end{equation}
where $O_{R}(t)$ is the complex amplitude of $O(t)$ and $\omega_{R}$ is its dominant response frequency. In Eq.~\eqref{eq:complex_amplitude}, $R(t)$ corresponds to the response amplitude of $O(t)$, and $\varphi(t)$ is the time-dependent absolute-time phase of the system.

To test the bit-flip operations for the DTC states, we consider the open Dicke model (ODM) described by the Lindblad master equation \cite{dimer_proposed_2007}
\begin{equation}
\label{eq:dm_master_eq}
\partial_{t} \hat{\rho} = -i \left[ \hat{H}/\hbar, \hat{\rho} \right] + \kappa \left( 2 \hat{a}\hat{\rho}\hat{a}^{\dagger} - \left\{ \hat{a}^{\dagger}\hat{a}, \hat{\rho} \right\}  \right),
\end{equation}
where the Hamiltonian is \cite{emary_chaos_2003, dimer_proposed_2007},
\begin{equation}
\label{eq:dm_hamiltonian}
\frac{\hat{H}}{\hbar} = \omega \hat{a}^{\dagger}{\hat{a}} + \omega_{0}\hat{S}^{z} + \frac{2\lambda(t)}{\sqrt{N}}\left( \hat{a}^{\dagger} + \hat{a} \right)\hat{S}^{x}. 
\end{equation}
The ODM, depicted in Fig.~\ref{fig:switch_diagram}(b), describes the interaction of $N$ two-level systems coupled to a single cavity mode that is connected to a Markovian bath \cite{emary_chaos_2003, dimer_proposed_2007}. The two-level systems are represented by the collective spin operators $\hat{S}^{x, y, z} = \sum_{i = 1}^{N} \hat{\sigma}_{i}^{x, y, z} / 2$, with $\sigma_{i}^{x, y, z}$ representing the $i$-th two-level system. Meanwhile, the cavity mode is represented by the bosonic creation (annihilation) operator $\hat{a}^{\dagger}$ ($\hat{a}$). The transition frequencies of the cavity and spins are $\omega$ and $\omega_{0}$, respectively. The time-dependent spin-cavity coupling is $\lambda(t) =\lambda_{0}f(t)$. Throughout this work, we consider $\omega = \omega_{0}$.

In the thermodynamic limit, $N \rightarrow \infty$, and static limit, $\lambda(t) \rightarrow \lambda_{0}$, the ODM has two equilibrium phases: the normal phase (NP) and the superradiant phase (SP) \cite{emary_chaos_2003, dimer_proposed_2007}. For coupling strengths less than a critical value, $\lambda_{0} < \lambda_{c}$, the system prefers the NP characterized by a fully polarized collective spin $S^{z} \equiv \langle \hat{S}_{z} \rangle$ at the $-z$ direction, and zero photon number. On the other hand, for $\lambda_{0} > \lambda_{c}$, the system enters the SP, which manifests as having a nonzero photon number and a nonzero $ S^{x} \equiv \langle \hat{S}^{x} \rangle$ component, the direction of which is chosen from the two degenerate steady states of the system \cite{dimer_proposed_2007}. The critical point separating the NP and the SP is $\lambda_{c} = \frac{1}{2} \sqrt{\kappa^{2} + \omega^{2}}.
$

When the ODM is periodically driven, with its initial state being either the NP or SP, the system can enter a DTC state. It manifests as a period-doubling dynamics with a $\varphi$ given by the phase of the oscillation of the two degenerate DTCs. For $\lambda_{0} < \lambda_{c}$, the ODM can be mapped onto a coupled PO with a resonance condition \cite{jara_jr_theory_2024, jager_dissipative_2023, chitra_dynamical_2015} 
\begin{equation}
\label{eq:dm_resonant_condition}
\omega_{r, \mathrm{DM}} = 2\omega_{-}, \quad A_{r, \mathrm{DM}} = \omega \sqrt{1 - \left( \frac{\lambda_{0}}{\lambda_{c}} \right)^{2}} \frac{\kappa}{\kappa^{2} + \omega^{2}},
\end{equation}
where $\omega_{-}$ is the lower polariton mode of the ODM \cite{jara_jr_theory_2024, dimer_proposed_2007},
\begin{equation}
\label{eq:lower_polariton}
\omega_{-}^{2} = \omega^{2} - \frac{\kappa^{2}}{4} - \omega \sqrt{ \left( \frac{\lambda_{0}}{\lambda_{c}} \right)^{2} \left( \omega^{2} + \kappa^{2} \right) - \kappa^{2}   }. 
\end{equation}
Note, however, that this mapping breaks down for the SP since it assumes negligible excitation of the two-level systems \cite{emary_chaos_2003}.

We investigate the bit-flip operation on the period-doubled and DTC states using a defect protocol based on a linear ramp of the phase of the drive. In particular, given a periodic drive with a time-dependent phase $\theta(t)$
\begin{equation}
	f(t) = 1 + A \sin\left[ \omega_{d}t + \theta(t) \right],
\end{equation}
we perform a bit-flip by applying the defect,
\begin{equation}
\label{eq:phase_ramp_protocol}
	\theta(t) = 
	\begin{cases}
	0; \quad \quad \quad  t_{i} \leq t \leq 0 \\
	2\pi t / T_{\delta}; \quad 0 < t < T_{\delta} \\	
	2\pi; \quad \quad T_{\delta} \leq t \leq t_{f} 
	\end{cases}	,
\end{equation}
where $T_{\delta}$ is the defect duration, and $t_{i}$ and $t_{f}$ are the initial and final time, respectively. Throughout this work, we consider the $x$-quadrature of the PO, $x = \sin(u)$, as its main order parameter, while we use $S^{x}$ for the ODM.

In the limit of $T_{\delta} \gg T_{d}$, the defect protocol in Eq.~\eqref{eq:phase_ramp_protocol} will always induce a bit flip on the period-doubled and DTC states since the absolute-time phase of both states adiabatically follow $\theta(t)$ when the ramp is sufficiently slow \cite{apffel_experimental_2024}. In particular, during the defect protocol, the driving frequency effectively becomes $\omega_{d}' = \omega_{d} + 2\pi / T_{\delta}$ for any $T_{\delta}$. Since the order parameter of both the PO and the ODM oscillates at half the driving frequency, we can infer that during the defect protocol for large $T_{\delta}$, 
\begin{equation}
O(t) \propto \cos\left( \frac{\omega_{d}}{2} t + \varphi_{0} + \frac{\pi}{T_{\delta}}t  \right), \quad 0 < t <T_{\delta},
\end{equation} 
where $\varphi_{0}$ is the absolute-time phase before the defect.
Thus at $t \geq T_{\delta}$, both systems acquire a phase shift of $\pi$, leading to a perfect bit-flip success rate for $T_{\delta} \gg T_{d}$. In the following section, we are interested in the response of the two systems and the robustness of the bit-flip operations when $T_{\delta}$ is close to $T_{d}$.

\section{Period-doubled and DTC state switching}\label{sec:parametric_state_swithing}

\subsection{Ideal limit without noise}\label{subsec:no_noise_switching}

Before we proceed, let us first establish the ideal noiseless limit for the PO and the ODM. From Eq.~\eqref{eq:po_noise_condition}, we can observe that the PO do not experience any stochastic noise when $\tilde{T} = 0$ or when $\gamma = 0$. As such, throughout this subsection, we will consider the zero-temperature limit and allow the dissipation $\gamma$ to be a control parameter. As for the ODM, as discussed in Appendix~\ref{sec:eom}, we reach the noiseless limit in the thermodynamic limit where the quantum fluctuations are suppressed. In this case, the dynamics of the system is described by the mean-field equations given in Appendix~\ref{sec:eom}.

We present in Figs.~\ref{fig:switch_diagram}(c)--\ref{fig:switch_diagram}(f) the zero-temperature bit-flip diagram of the PO for increasing $\gamma$. The bit-flip diagram identifies the parameter regimes where we can successfully flip a bit encoded in the period-doubled state. Here, $\delta A = A - A_{r}$ corresponds to the detuning of the driving amplitude from $A_{r}$. We also show in Figs.~\ref{fig:switch_diagram}(g)-\ref{fig:switch_diagram}(j) the same set of bit-flip diagrams for the ODM for increasing $\kappa$ and $\lambda_{0} = 0.9\lambda_{c}$ in the thermodynamic limit.

To determine whether we can switch a period-doubled or DTC state at a given $\delta A$ and $T_{\delta}$, we first initialize the systems in their stable fixed points at $t_{i} < 0$. For the PO, we consider the initial state
\begin{equation}
\label{eq:dpp_initial_state}
u_{0} = 10^{-4}, \quad \dot{u}_{0} = 0,
\end{equation}
while for the ODM, we initialize in the steady state corresponding to the NP:
\begin{equation}
\label{eq:dm_initial_state}
a_{0} = \epsilon \sqrt{N}, \quad S^{x}_{0} = \epsilon \frac{N}{2}, \quad  S^{y}_{0} = 0, \quad S^{z}_{0} = -\frac{N}{2} \sqrt{1 - \epsilon^{2}},
\end{equation}
where $a \equiv \langle \hat{a} \rangle$ and $S^{x, y, z} = \langle \hat{S}^{x, y, z}\rangle $ with $\epsilon = 10^{-6}$. We then apply a periodic drive to push the PO (ODM) into a period-doubled (DTC) state for 100 driving cycles. At $t = 0$, we finally apply the defect protocol in Eq.~\eqref{eq:phase_ramp_protocol} until $t = T_{\delta}$, as exemplified in Fig.~\ref{fig:schematics}(b). We then allow the system to relax back to a new period-doubled (DTC) state before obtaining the difference between the steady-state values of the absolute-time phase before and after the defect, $\triangle \varphi$. If $\triangle \varphi = \pi$, we have a successful bit-flip, while we have no bit-flip when $\triangle \varphi = 0$. Note that in evaluating $\triangle \varphi$, we enforced the periodic boundary of the absolute-time phase such that $\varphi \in [-\pi, \pi)$.

While the bit-flip diagrams of the PO and the ODM in Fig.~\ref{fig:switch_diagram} appear to be quantitatively distinct from each other, there is a clear qualitative trend for varying dissipation strengths. For weak dissipation, for instance, the bit-flip diagram exhibits complex, fractal-like structures. These fractal structures, as discussed in Appendix~\ref{sec:fractals}, emerge due to the transient irregular dynamics induced by the defect protocol. These transient irregular dynamics becomes suppressed, however, by strong dissipation, resulting to the loss of fractal structures at large dissipation strengths. This behavior suggests that in the zero-temperature limit and the thermodynamic limit for the PO and ODM, respectively, bit-flip operations become more robust when the dissipation is strong, in that perturbations in $A$ and $T_{\delta}$ will not significantly affect the success rate of the bit-flip between the states. This is in contrast to weak dissipation, where there is more uncertainty in the bit-flip success due to the fractal structures in the bit-flip diagram.

Given that strong dissipation allows for a more robust bit-flip between period-doubled and DTC states, we will determine in the next subsection whether the bit-flip operation remains robust even in the presence of thermal and quantum fluctuations for the PO and ODM, respectively.

\subsection{Effects of thermal and quantum noise}\label{subsec:with_noise_switching}

We explore the robustness of the bit-flip operation of period-doubled states against thermal noise by numerically integrating the Langevin equation in Eq.~\eqref{eq:stochastic_pendulum} using a predictor-corrector method, with a time step of $\Omega \triangle t = 0.01$. The initial state is the same as in the zero-temperature limit given by Eq.~\eqref{eq:dpp_initial_state}. We find that high temperatures destroy the coherence of the parametric oscillation in the long-time limit (see Appendix \ref{subsec:stable_dtc_under_noise} for details), and as such, we only consider a range of temperature from $\tilde{T} = 10^{-6}$ to $\tilde{T} = 2 \times 10^{-4}$.

As for the DTCs in the ODM, a way to account the quantum fluctuations is to explicitly solve the Lindbland master equation in Eq.~\eqref{eq:dm_master_eq} to obtain the density matrix $\hat{\rho}(t)$ of the system. While this gives the exact quantum dynamics of the ODM, this method is computationally expensive for large $N$ since the Hilbert space of the system grows exponentially with $N$. To circumvent this problem, we employ an approximation called the truncated Wigner approximation (TWA) and its counterpart for spin systems, the discrete truncated Wigner approximation (DTWA) to solve the dynamics of the system. The TWA and DTWA work by evolving an ensemble of trajectories with initial values sampled from the Wigner distributions of the spin and cavity modes, and extracting the statistics of an observable from the ensemble \cite{polkovnikov_phase_2010, huber_realistic_2022, schachenmayer_many-body_2015}. In particular, for both the TWA and DTWA, we treat the cavity mode $\hat{a}$ as a complex variable $a(t)$ with a vacuum state as its initial state. This can be represented as a complex Gaussian variable $a(t_{i}) = \left(\zeta_{\mathrm{R}} + i \zeta_{\mathrm{I}} \right) / 2$ satisfying the conditions $\langle \zeta_{i} \rangle = 0$ and $\langle \zeta_{i}\zeta_{j} \rangle = \delta_{i, j}$ for $i, j \in \left\{ \mathrm{R}, \mathrm{I} \right\}$ \cite{olsen_numerical_2009}. As for the spin component, in TWA, we treat the collective spin operators $\hat{S}^{x, y, z}$ as real variables with initial states given by Eq.~\eqref{eq:dm_initial_state}. Whereas for the DTWA, we treat each two-level system $\hat{\sigma}_{i}^{x, y, z}$ as separate real variables, with $S^{x, y, z} = \sum_{i = 1}^{N}\sigma_{i}^{z, y, z} / 2$. In this case, we initialize the $z$-component of the individual spins at $\sigma^{z}_{i} = -1$, while we sample the $x$ and $y$ components of the spins from the set $(\sigma_{i}^{x}, \sigma_{i}^{y}) \in \left\{ (1, 1), (1, -1), (-1, 1), (-1, -1)   \right\}$ with equal probability. Thus, the DTWA captures the quantum noise contributed by each two-level system.

Both the TWA and DTWA account for the temporal noise neglected in the mean-field equations by including a stochastic term in the system's equations of motion. In particular, given that the ODM is connected to a Markovian bath, the dynamics of the cavity and spin components can be described instead by the Heisenberg-Langevin equation \cite{ritsch_cold_2013},
\begin{subequations}
\label{eq:heisenberg_langevin}
\begin{equation}
\partial_{t} \hat{a} = i\left[ \hat{H}/\hbar, \hat{a} \right] - \kappa \hat{a} + \hat{\xi}(t),
\end{equation}
\begin{equation}
\partial_{t}\hat{S}^{x, y, z} = i \left[\hat{H} /\hbar, \hat{S}^{x, y, z} \right], 
\end{equation}
\begin{equation}
 \partial_{t}\hat{\sigma}_{i}^{x, y, z} = i \left[\hat{H} / \hbar, \hat{\sigma}_{i}^{x, y, z} \right],
\end{equation}
\end{subequations}
where $\hat{\xi}(t)$ is a Gaussian noise operator satisfying the conditions, 
\begin{equation}
\left< \hat{\xi}(t) \right> = 0, \quad
\left< \hat{\xi}(t)\hat{\xi}(t') \right> = \kappa \delta\left(t-t'\right).
\end{equation}
The noise term $\hat{\xi}$ can be viewed as fluctuations due to the photons entering the system from the Markovian bath as shown in Fig.~\ref{fig:switch_diagram}(b). Mathematically, $\hat{\xi}(t)$ preserves the commutation relation of the cavity mode after evaluating the commutations in Eq.~\eqref{eq:heisenberg_langevin} and replacing the operators $\hat{a}$, $\hat{S}^{x, y, z}$, and $\sigma^{x, y, z}$, with the complex variables $a$, $S^{x, y, z}$, and $\sigma^{x, y, z}$, respectively. Following this procedure results to the stochastic equations of motions shown in Appendix~\ref{sec:eom}. With these equations of motion, we can calculate the system's dynamics for an ensemble of initial states and obtain the statistics of the quantities we are interested in. For both the TWA and DTWA, we solve the equations of motion using the predictor-corrector method with a time step of $\omega \triangle t = 0.01$

\begin{figure}
\centering
\includegraphics[scale=0.45]{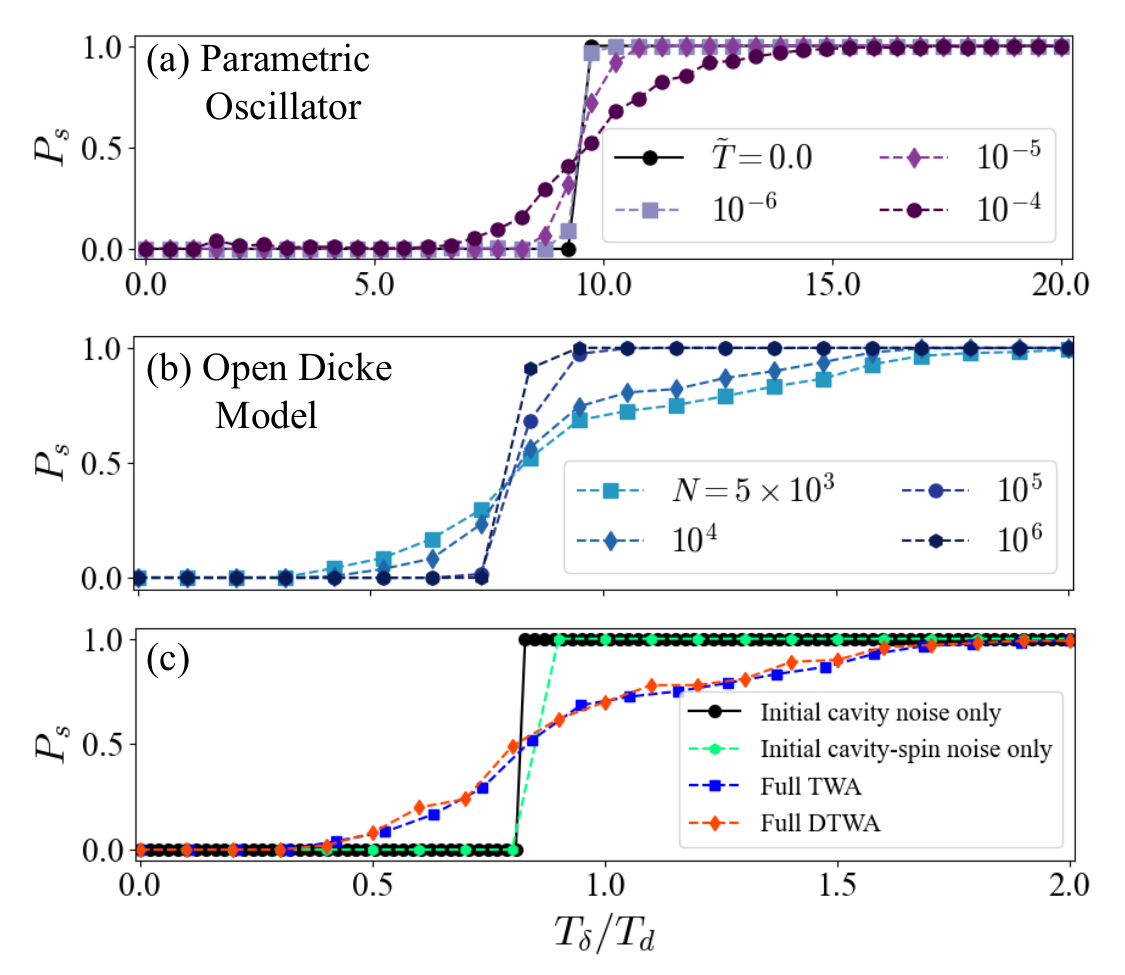}
\caption{(a) Switching probability of the resonantly driven PO as a function of $T_{\delta}$ for $\gamma = 0.1\Omega$, $\delta A = 0.4$, and different values of $\tilde{T}$. (b), (c) Switching probability of the resonantly driven ODM for (b) different $N$ and (c) different set of stochastic noise for $N = 5\times 10^{3}$. For (b), the dynamics are obtained using the TWA. The remaining parameters are $\lambda_{0} = 0.9 \lambda_{c}$, $\kappa = 1.0\omega$, and $\delta A = 0.1$. For both systems, we considered $1000$ trajectories.}
\label{fig:switch_probability}
\end{figure}

We present in Fig.~\ref{fig:switch_probability}(a) the success probability of a bit flip between period-doubled states, $P_{s}$, as a function of $T_{\delta}$ for $\tilde{T}\neq 0$ and $1000$ trajectories. We can observe that for very small $\tilde{T}$, $P_{s}$ has a discontinuous transition from $P_{s} = 0$ to $1$ at $T_{\delta, c}\approx 10 T_{d}$, akin to those observed in Ref. \cite{apffel_experimental_2024}. As we increase the temperature, however, the thermal fluctuations soften the discontinuity into a crossover. This allows the system to have a nonzero probability to switch between period-doubled states at lower $T_{\delta}$. We can see a similar effect of quantum noise on the DTCs for finite $N$, as depicted in Fig.~\ref{fig:switch_probability}(b). As expected, the transition between unsuccessful and successful bit-flips becomes sharper as $N$ increases and approaches the shape of that in the thermodynamic limit, $N \rightarrow \infty$. This can be understood from the inverse proportionality between the temporal noise strength and the particle number as highlighted in Eq.~\eqref{eq:rescaled_open_dm_eom}.

The quantum noise on the ODM originates from two sources: (i) initial quantum fluctuations and (ii) temporal fluctuations due to the dissipation. As such, it is natural to ask how these two types of fluctuations contribute to the crossover-like behavior of $P_{s}$. To this end, we investigate the $P_{s}$ of the ODM when there is only initial noise in the cavity and spin degrees of freedom. We present this in Fig.~\ref{fig:switch_probability}(c) together with the $P_{s}$ obtained using the TWA and the DTWA. Notably, when only initial noise is present on the cavity and individual spins, we observe the discontinuous transition seen in the large-$N$ limit, which is in contrast to the crossover-like feature of $P_{s}$ in the TWA and DTWA. We attribute this to dissipation removing any effects of the initial noise on the dynamics in the absence of temporal fluctuations, leading to an effectively mean-field dynamics on a single-trajectory level, and thus the mean-field characteristics of $P_{s}$.

\begin{figure}
\centering
\includegraphics[scale=0.5]{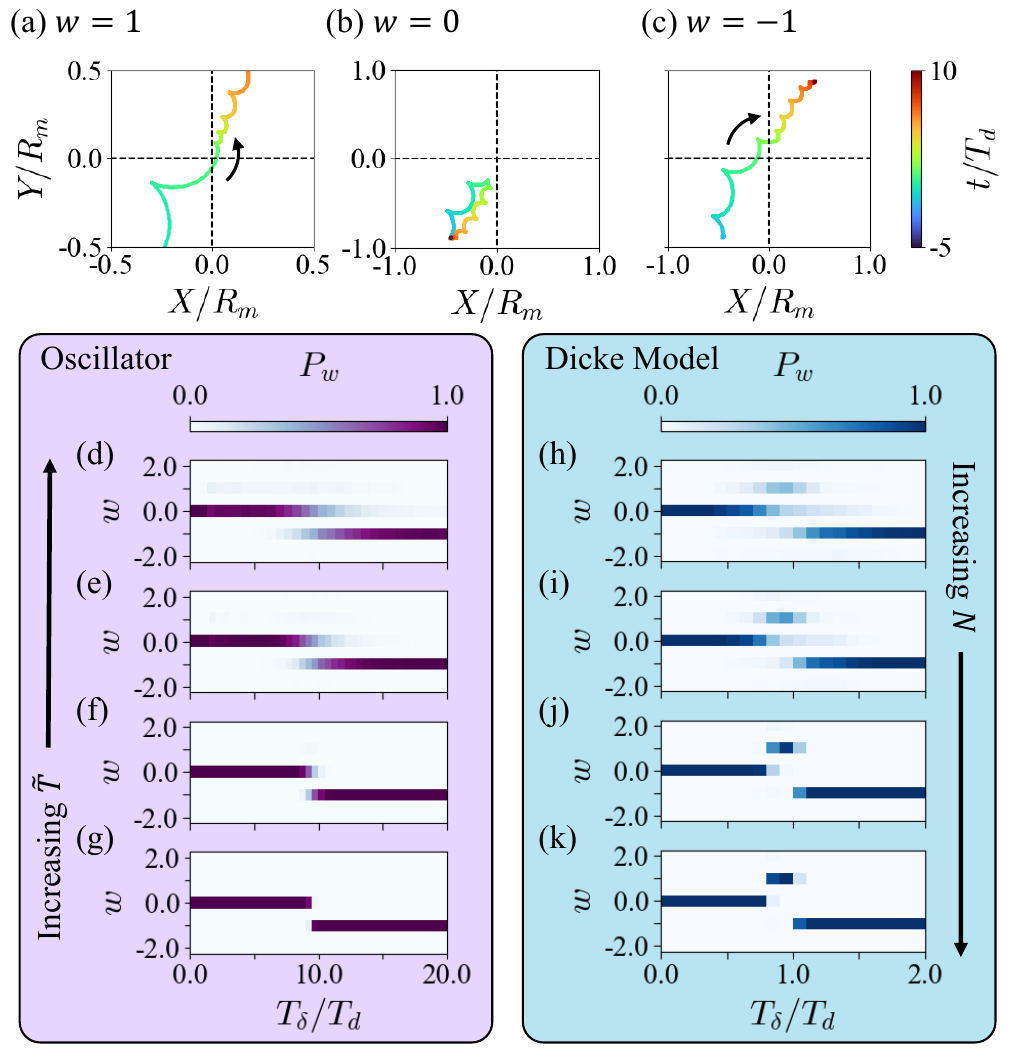}
\caption{(a)--(c) Exemplary dynamics of the complex response amplitude of $S^{x}/N$ for (a) $w = 1$, (b) $w = 0$, and (c) $w = -1$. The corresponding defect durations are (a) $T_{\delta} = 0.95 T_{d}$, (b) $T_{\delta} = 0.5T_{d}$, and (c) $T_{\delta} = 1.5$, respectively. (d)--(g) Half-winding number distribution of the resonantly-driven PO for (d) $\tilde{T} = 2\times 10^{-4}$, (e) $\tilde{T} = 10^{-4}$, (f) $\tilde{T} = 10^{-5}$, and (g) $\tilde{T} = 10^{-6}$. (h)--(k) Half-winding number distribution of the ODM at resonance for (h) $N = 5\times 10 ^{3}$, (i) $N = 10^{4}$, (j) $N = 10^{5}$, (k) $N = 10^{6}$. The remaining parameters for the PO are $\gamma = 0.1\Omega$, and $\delta A = 0.4$, while the remaining parameters for the ODM are $\lambda_{0} = 0.9\lambda_{c}$, $\kappa = 1.0 \omega$, and $\delta A = 0.1$.}
\label{fig:winding_probability}
\end{figure}

In Fig.~\ref{fig:switch_probability}(b), unlike the crossover in Fig.~\ref{fig:switch_probability}(a), the $P_{s}$ of the ODM for $N = 5\times 10 ^{3}$ and $10^{4}$ exhibits a plateau within the interval $T_{\delta} / T_{d} \in \left[1.0, 1.5 \right]$ before approaching $P_{s} = 1$ for large $T_{\delta}$. This effect is more apparent in Fig.~\ref{fig:switch_probability}(c), wherein the $P_{s}$ approaches a plateau at $P_{s} \approx 0.75$. To better understand the appearance of such plateaus, we further characterize the switching dynamics of the period-doubled and DTC states using the half winding number $w$, introduced in Ref.~\cite{apffel_experimental_2024} to characterize the dynamics of a period-doubled state during the defect. It is defined as \cite{apffel_experimental_2024, simula_topological_2024}, 
\begin{equation}
\label{eq:half_winding_number}
w = \frac{1}{\pi} \int_{0}^{\infty} \frac{\partial \varphi(t')}{\partial t'} dt' = \frac{1}{\pi} \left[ \lim_{t \rightarrow \infty} \varphi(t) - \varphi(0) \right],
\end{equation}
and it quantifies the number of times the $O_{R}(t)$ performs a half-rotation around the origin of the complex plane. To illustrate this for a DTC during a defect protocol, we present in Figs.~\ref{fig:winding_probability}(a)--\ref{fig:winding_probability}(c) the exemplary dynamics of $X = \mathrm{Re}[O_{R}]$ and $Y = \mathrm{Im}[O_{R}]$. Here, we rescale the axes by the maximum value of $|R(t)|$, $R_{m}$. Similar to the results in Ref.~\cite{apffel_experimental_2024}, and Figs.~\ref{fig:schematics}(d), \ref{fig:schematics}(f), and \ref{fig:schematics}(h), during the defect protocol, the trajectory of the $O_{R}$ approaches the origin. The half-winding number depends on whether $O_{R}$ rotates around the origin or not. If $O_{R}$ avoids the origin, then $w=0$, and thus there is no bit-flip. On the other hand, if $O_{R}$ does a single half counter-clockwise rotation around the origin, then $w = +1$, meaning the period-doubled or DTC state switches to its symmetry-broken partner. The same thing occurs when $w=-1$, except that $O_{R}$ now rotates in a clockwise manner. Note that we only get a bit-flip when $w = 2n + 1$, with $n \in \mathbb{Z}$, since an even-valued $w$ implies a complete rotation around the origin and that the system returns to its original state. Note that our convention for obtaining $\varphi(t)$ using Eq.~\eqref{eq:complex_amplitude} results to a half-winding number of $w = -1$ in the limit of $T_{\delta}\gg T_{d}$ for both the PO and the ODM.

We now show in Figs.~\ref{fig:winding_probability}(d)--\ref{fig:winding_probability}(g) the probability distribution $P_{w}$ of $w$ as a function of $T_{\delta}$ and $\tilde{T}$ for the thermal PO. Note that we have used the same parameters considered in Fig.~\ref{fig:switch_probability}(a). We can see that the $P_{w}$ is localized in a small subset of $w$ when $\tilde{T}\ll 1$. Specifically, for $T_{\delta} < T_{\delta, c}$, $w = 0$, and thus no bit flip occurs. Whereas for $T_{\delta} > T_{\delta, c}$, $w = -1$, implying that the period-doubled state switches as a result of the defect protocol. We can also observe that $w$ is discontinuous exactly at $T_{\delta} = T_{\delta, c}$, consistent with the behavior observed in Ref.~\cite{apffel_experimental_2024}. However, as we increase $\tilde{T}$, the discontinuity at $T_{\delta, c}$ becomes blurry as $P_{w}$ spreads out such that the PO can either have $w = 0$ or $-1$ near $T_{\delta, c}$, i.e., it is now uncertain if the bit-flip operation will work or not. This behavior is consistent with the crossover observed in Fig.~\ref{fig:switch_probability}(a). Note that in the case of $\tilde{T} = 2\times 10 ^{-4}$, as shown in Fig.~\ref{fig:winding_probability}(d), the faint $w = +1$ branch in the $P_{w}$ for $T_{\delta} < T_{\delta, c}$ is due to thermally-activated switching, in which due to the finite coherence time of the oscillations, the thermal noise can become strong enough to flip a period-doubled state \cite{marthaler_switching_2006}.

The situation is different in the case of the $P_{w}$ of the ODM, as shown in Figs.~\ref{fig:winding_probability}(h)--\ref{fig:winding_probability}(k). While we still observe qualitatively similar behavior of the distribution as in the PO, such as the localization of $P_{w}$ into one value of $w$ as $N$ increases, the $P_{w}$ of the ODM also reveals a $w = +1$ branch from $T_{\delta} \approx 0.9 T_{d}$ to $1.0 T_{d}$. In Fig.~\ref{fig:winding_probability}(k), we show that when $N$ is sufficiently large, the $P_{w}$ in these regions of $T_{\delta}$ becomes more concentrated into $w = +1$, explaining the discontinuous transition at $T_{\delta} \approx 0.9 T_{d}$ shown in Fig.~\ref{fig:switch_probability}(b). As we decrease $N$, however, the system obtains nonzero probability to either have a half-winding number of $w = \pm  1$ or $0$, thus resulting in the plateau observed in Figs.~\ref{fig:switch_probability}(b) and \ref{fig:switch_probability}(c). Note that while we do not observe any appearance of plateaus in the thermal PO for the parameters considered here, we still expect this behavior to appear when the system has a transition between $w = -1$ and $+1$, and the range of $T_{\delta}$ in which one of the branches exists is small enough for the system to have nonzero probability to get either $w = 0$ or $\pm 1$.

Our results imply two things about the robustness of the bit-flip operation beyond the weak-dissipation and noiseless limit. First, in the noiseless limit, strong dissipation enhances the robustness of the operation against perturbations in the driving and defect parameters. This leads to a wider range of $A$ and $T_{\delta}$ wherein controlled switching between period-doubled and DTC states can be observed. Second, in the presence of fluctuation for fixed dissipation, not only does the bit-flip operation remains robust for large $T_{\delta}$, but the noise also enhances the switching probability for $T_{\delta} < T_{\delta, c}$. Thus, we demonstrate that the phase ramp as a defect is a robust method for performing bit-flips between the period-doubled states of thermal PO and the DTC states of finite-sized ODM for $\lambda_{0} < \lambda_{c}$.

Given that the DTC states of the ODM for $\lambda_{0} < \lambda_{c}$ can be mapped onto a period-doubled state of a coupled oscillator model \cite{jara_jr_theory_2024}, we now explore in the next subsection whether we can still switch DTC states using the defect protocol in Eq.~\eqref{eq:phase_ramp_protocol} when $\lambda_{0}>\lambda_{c}$, in which the mapping of the ODM onto the PO breaks down.

\subsection{Beyond parametric oscillator models}\label{subsec:beyond_oscillators}

\begin{figure}
\centering
\includegraphics[scale=0.68]{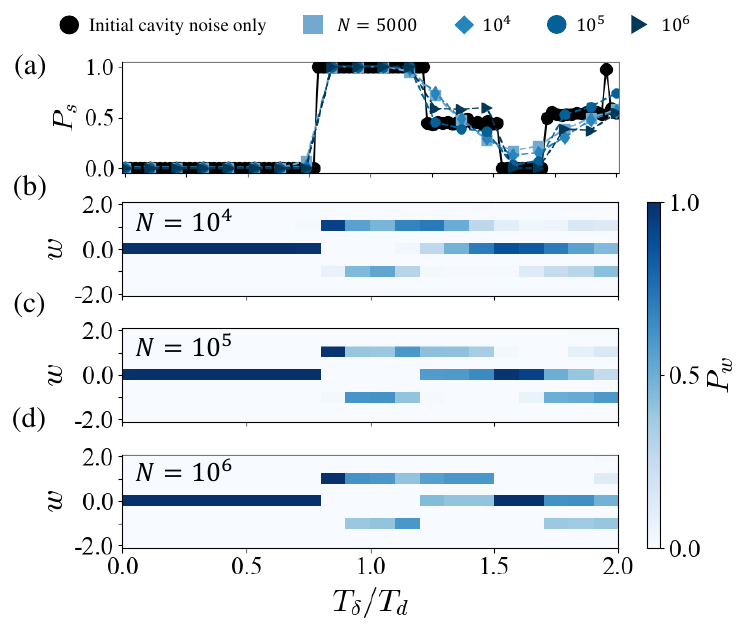}
\caption{(a) Switching probability of the ODM for different particle number and for the parameters $\lambda_{0} = 1.1 \lambda_{c}$, $\kappa = 1.0\omega$, $\omega_{d} = 0.8\omega$, and $A = 0.55$. (b)-(d) Half-winding number distribution of the ODM for (b) $N = 10^{4}$, (c) $N = 10^{5}$, (d) $N = 10^{6}$. The same set of parameters in (a) is used to construct the histograms in (b)--(e).}
\label{fig:sr_winding_stats}
\end{figure}

We present in Fig.~\ref{fig:sr_winding_stats}(a) the bit-flip probability of the DTC states in the $\lambda_{0} > \lambda_{c}$ regime. Notice that the $P_{s}$ has a more complex behavior, with its most striking feature being the discontinuity at $T_{\delta, c} \approx 0.8T_{d}$ that persists even for small $N$. This is in contrast to the behavior observed so far with the thermal PO and DTC states in the NP in Figs.~\ref{fig:switch_probability}(a) and \ref{fig:switch_probability}(b), respectively. As we further increase $T_{\delta}$, however, $P_{s}$ drops to a lower value. In particular, when we neglect the temporal fluctuations and only consider the initial noise, $P_{s}\approx 0.5$ within the range of $T_{\delta} / T_{d} \in \left[1.2, 1.5\right]$ and $T_{\delta} / T_{d} \in \left[1.7, 2.0 \right)$, while it goes to zero for $T_{\delta} / T_{d} \in \left(1.5, 1.7 \right)$. Unlike the discontinuous transition in $T_{\delta} = 0.8 T_{d}$ however, this feature does not persist for small $N$, and instead becomes smoothened out similar to the crossovers observed in Figs.~\ref{fig:switch_probability}(a) and \ref{fig:switch_probability}(b).

From the distribution of $w$ shown in Figs.~\ref{fig:sr_winding_stats}(b)--\ref{fig:sr_winding_stats}(d), we see that at $T_{\delta, c} = 0.8 T_{d}$, the distribution jumps from $w = 0$ to $+1$ and then splits into $w = +1$ and $-1$ as we further increase $T_{\delta}$. This splitting of the $P_{w}$ persists even for large values of $N$, which, again, is in contrast to the thermal PO and the DTC states for $\lambda_{0} < \lambda_{c}$, where the distribution only becomes localized at one value of $w$. This splitting persists even for large $T_{\delta}$, with $P_{w}$ localizing at $w = 0$ and $+1$ in the defect time interval $T_{\delta} / T_{d} \in \left[1.2 , 1.5 \right]$, and at $w = 0$ and $-1$ for $T_{\delta} / T_{d} \in \left[1.7 , 2.0 \right)$.

\begin{figure}
\centering
\includegraphics[scale=0.44]{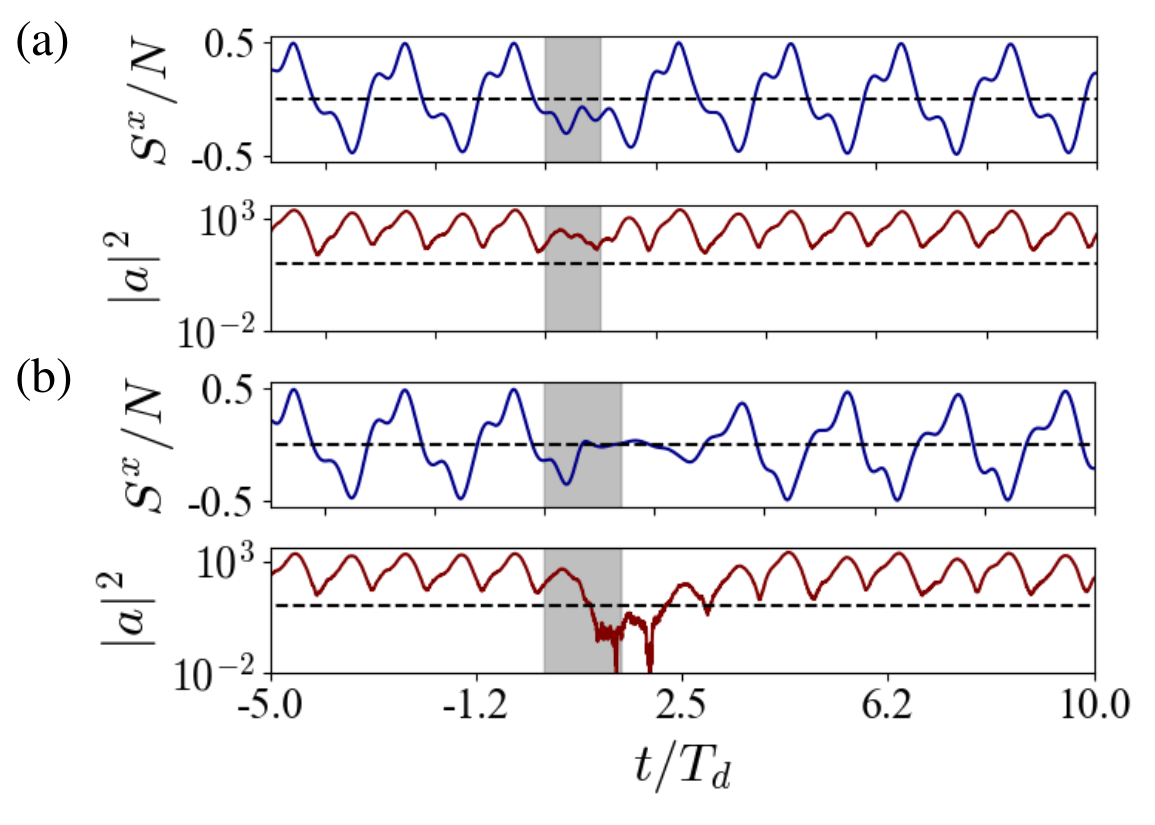}
\caption{Single-trajectory dynamics of the ODM for (a) $T_{\delta} = 1.0 T_{d}$ and (b) $T_{\delta}=1.4 T_{\delta}$, with $\lambda_{0} = 1.1\lambda_{c}$, $\kappa = 1.0\omega$, $\omega_{d} = 0.8\omega$, $A = 0.5$, and $N = 5000$ particles. Top panels show the dynamics of $S^{x}$, while the bottom panels shows the photon number, $|a|^{2}$. The gray regions denote the time interval at which the defect protocol is on, while the solid dashed lines marks $S^{x} = 0$ and $|a|^{2}=10$ in the top and bottom panels, respectively. }
\label{fig:sr_switch_dynamics}
\end{figure}

To understand why these features appear for $\lambda_{0} > \lambda_{c}$, we study the dynamics of $S^{x}$ and the photon number, $|a|^{2}$ during the defect protocol. We present in Fig.~\ref{fig:sr_switch_dynamics}(a) the dynamics of these two quantities for $N = 5000$ when $T_{\delta} = T_{d}$, which is close to the discontinuous transition point in $T_{\delta} \approx 0.8 T_{d}$. Notice that unlike the DTC states for $\lambda_{0} < \lambda_{c}$, $S^{x}$ does not go to zero throughout the defect. Instead, it retains a nonzero value, implying that as soon as the defect protocol is turned on, the system reverts to the SP. This is further highlighted in the bottom panel of Fig.~\ref{fig:sr_switch_dynamics}(a), which depicts that the photon number during the defect is larger than the threshold $|a|^{2} = 10$ around which we expect the effects of the quantum fluctuations to become relevant. This explains not only the robustness of the bit-flip operation for $T_{\delta}$ close to $T_{\delta, c} = 0.8 T_{d}$, but also why the discontinuous transition at this critical point survives even for small $N$. We demonstrate in Fig.~\ref{fig:sr_switch_dynamics}(b), however, that as soon as we consider larger $T_{\delta}$, the defect protocol can push the system into a light-induced NP, in which both $S^{x}$ and $|a|^{2}$ approach zero \cite{cosme_dynamical_2018, georges_light-induced_2018}. This results in the system becoming significantly affected by the quantum noise, reducing the overall robustness of the bit-flip operation.

Next, to explain the splitting of the $P_{w}$ in Fig.~\ref{fig:sr_winding_stats}, let us again consider the dynamics shown in Fig.~\ref{fig:sr_switch_dynamics}(a). We note that depending on the state of the system before the defect protocol at $t = 0$ for $T_{\delta} / T_{d} \in \left[0.8 , 1.2  \right)$, $S^{x}$ can either oscillate around a positive or a negative value during the defect protocol, consistent with the $\mathbb{Z}_{2}$-symmetry breaking nature of the SP. Due to this additional degree of freedom, the complex amplitude of $S^{x}$ can revolve around $A_{R} = 0$ either clockwise or counter-clockwise, leading to the splitting of the half-winding distribution to $w = +1$ and $-1$. For ranges of defect duration given by $T_{\delta} / T_{d} \in \left[1.2 , 1.5 \right]$, and  $T_{\delta} / T_{d} \in \left[1.7, 2.0 \right)$, both the $S^{x}$ and $|a|^{2}$ are close to zero. This leads to quantum fluctuations removing the initial memory of the system, and thus, at $t > T_{\delta}$, the fluctuations can either push the system to its original oscillation pattern or send it to a new DTC state. This then explains the splitting of the half-winding distribution at $w = 0$ and $\pm 1$, depending on the value of $T_{\delta}$. Note that this bit-flip mechanism is different from the cases of the period-doubled states and DTC states for $\lambda_{0} < \lambda_{c}$, where the oscillation becomes macroscopic as soon as $T_{\delta} > T_{\delta, c}$, as shown for instance in Fig.~\ref{fig:schematics}(h), which then restricts $w$ to a single value.

To summarize our results in this section, we demonstrate the robustness of bit-flip operations on period-doubled and DTC states using the defect protocol in Eq.~\eqref{eq:phase_ramp_protocol} even for strong dissipation and noise strength, which is of thermal nature for the PO and quantum origin for the ODM. Moreover, we have also shown that we can still switch DTC states even at parameter regimes where the system cannot be mapped onto a standard PO. In particular, depending on how the system responds during the defect protocol, its transient state may even protect it from quantum fluctuations, leading to a more robust bit-flip operation. Due to the possibility of quenching the system into a new dynamical phase during the defect, we will now consider in the following section a generalization of the defect protocol that may allow us to fine-tune the system's state during the defect to obtain a more precise bit-flip operation.

\section{Generalized defect protocol}\label{sec:modified_defects}

\begin{figure}
\centering
\includegraphics[scale=0.71]{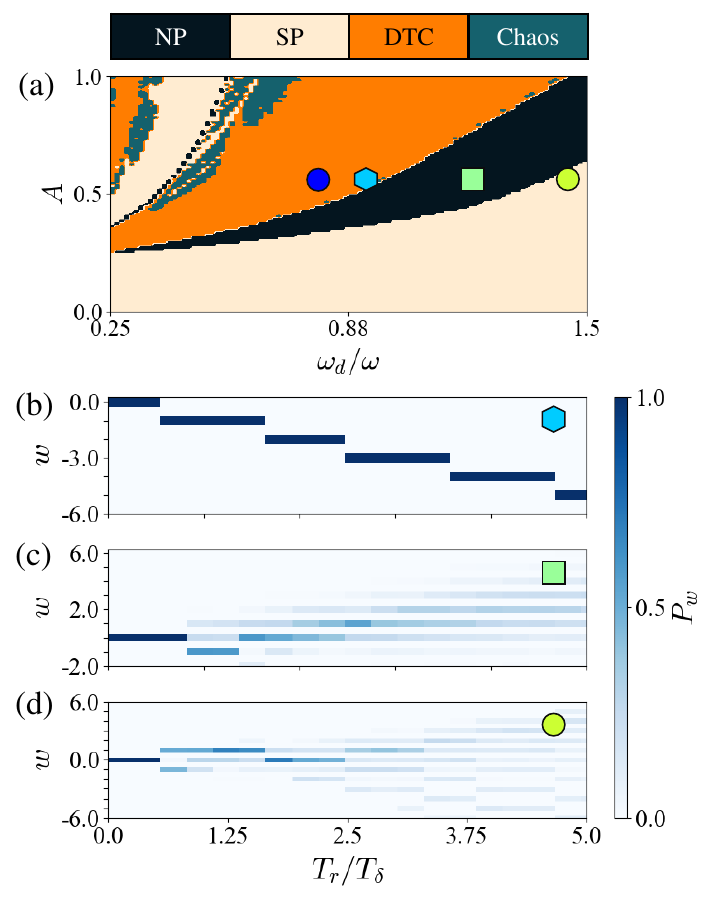}
\caption{(a) Phase diagram of the ODM for $\lambda_{0} = 1.1 \lambda_{c}$, and $\kappa = 1.0\omega$ in the absence of any defects. (b), (c) Half-winding number distribution of the ODM under the quenched-frequency protocol for (b) $\omega_{d}' = 0.9\omega$, (c) $\omega_{d}' = 1.2\omega$, and (d) $\omega_{d}' = 1.45\omega$. The dark circles in (a) marks the initial choice of $\omega_{d} = 0.8 \omega$ and the remaining markers denote the driving frequencies during the defect protocol $\omega_{d}'$ used in (b)--(d). The remaining driving parameters are $A = 0.55$, and $N = 5000$.  }
\label{fig:quenched_frequency}
\end{figure}

As discussed in Sec.~\ref{subsec:beyond_oscillators}, by applying the defect protocol according to Eq.~\eqref{eq:phase_ramp_protocol} on a DTC state for $\lambda_{0} > \lambda_{c}$, the ODM can be pushed into a new phase while the defect is switched on. To explain this behavior, note that Eq.~\eqref{eq:phase_ramp_protocol} can also be viewed as a sudden quench of the driving frequency from $\omega_{d}$ to $\omega_{d}' = \omega_{d} + 2\pi / T_{\delta}$. It is then fixed at $\omega_{d}'$ for some duration $T_{\delta}$ until the driving frequency is reverted to $\omega_{d}$ at $t > T_{\delta}$. In Fig.~\ref{fig:quenched_frequency}(a), we show that the dynamical phase of the ODM for $\lambda_{0} > \lambda_{c}$ depends on $\omega_{d}$. As such, applying a defect effectively pushes the system onto a new phase if $T_{\delta}$ is sufficiently large for the system to relax to the stable phase at a given $\omega_{d}'$. In the following, we exploit this driving-controlled quench of the dynamical phase to switch the bit encoded in the DTC states.

We now consider a modification of the defect protocol, which corresponds to quenching the driving frequency
\begin{equation}
\label{eq:quenched_frequency}
\lambda(t) = 
\begin{cases}
\lambda_{0}\left[ 1 + A\sin(\omega_{d} t) \right], \quad \quad t \leq 0 \\
\lambda_{0}\left[ 1 + A\sin\left( \omega_{d}'t \right) \right], \quad \quad 0 < t < T_{r} \\
\lambda_{0}\left[ 1 + A\sin(\omega_{d} t) \right], \quad \quad t \geq T_{r}
\end{cases},
\end{equation}
where we set $\omega_{d}'= \omega_{d} +  2\pi / T_{\delta}$ to be a constant value, while we treat the duration of the defect $T_{r}$ as a separate tuning parameter. Unlike in Eq.~\eqref{eq:phase_ramp_protocol}, the protocol in Eq.~\eqref{eq:quenched_frequency} is only continuous for $T_{r} = n T_{\delta}$, with $n \in \mathbb{Z}$, otherwise the phase of the drive jumps from $2\pi T_{r} / T_{\delta}$ to zero after the defect protocol. We consider a continuous version of this protocol in Appendix~\ref{sec:defect_with_phase_error}.

We present in Figs.~\ref{fig:quenched_frequency}(b)--\ref{fig:quenched_frequency}(d) the $P_{w}$ of the ODM for $\lambda_{0} > \lambda_{c}$ and varying $\omega_{d}'$ representing quenches into three different states: a quench to a DTC with a different response frequency; a quench to the light-induced NP; and a quench to the SP, as shown in Figs.~\ref{fig:quenched_frequency}(b)--\ref{fig:quenched_frequency}(d), respectively. We observe that among the three scenarios, we see a more localized distribution when we quench our system to another DTC phase with a different response frequency. In this case, the $P_{w}$ alternates between even and odd $w$, forming a staircase localized only at integer values of $w$. This behavior implies that the success of the bit-flip operation for this protocol only depends on $T_{r}$, making it a timing-based method for performing bit-flips.

For quenches into the light-induced NP, as exemplified in Fig.~\ref{fig:quenched_frequency}(c), we observe a spreading of the half-winding distribution, which becomes more prominent as we increase $T_{r}$. This can be attributed to the thermalization of the system as it relaxes into the light-induced NP during the defect protocol. As a result, the success of the bit-flip operation will highly depend on its state after $t = T_{r}$, which is then set by the quantum fluctuations present at that particular time. Note that while the $P_{w}$ spreads for the light-induced NP case, it remains localized in integer values of $w$ due to the particular form of the protocol in Eq.~\eqref{eq:quenched_frequency}, in which the driving phase reverts to zero at $t > T_{r}$. This restricts the dynamics after the defect protocol to the original degenerate DTC states, thereby leading to a difference between the initial and final absolute phase of $\triangle\varphi \equiv \varphi(t\rightarrow \infty) - \varphi(0) = \pi w$, with $w \in \mathbb{Z}$.

Finally, for the case of the quench into the SP, we show in Fig.~\ref{fig:quenched_frequency}(d) that the $P_{w}$ spreads similarly to that for the light-induced NP case. We can attribute this behavior to our particular choice of $\omega_{d}'$, wherein as shown in Fig.~\ref{fig:quenched_frequency}(a), our chosen $\omega_{d}'$ is close to the critical line separating the SP and the light-induced NP. As a result, the photon number during the defect protocol is not macroscopic enough to overcome quantum fluctuations, leading to a delocalized $P_{w}$.

\begin{figure}
\centering
\includegraphics[scale=0.51]{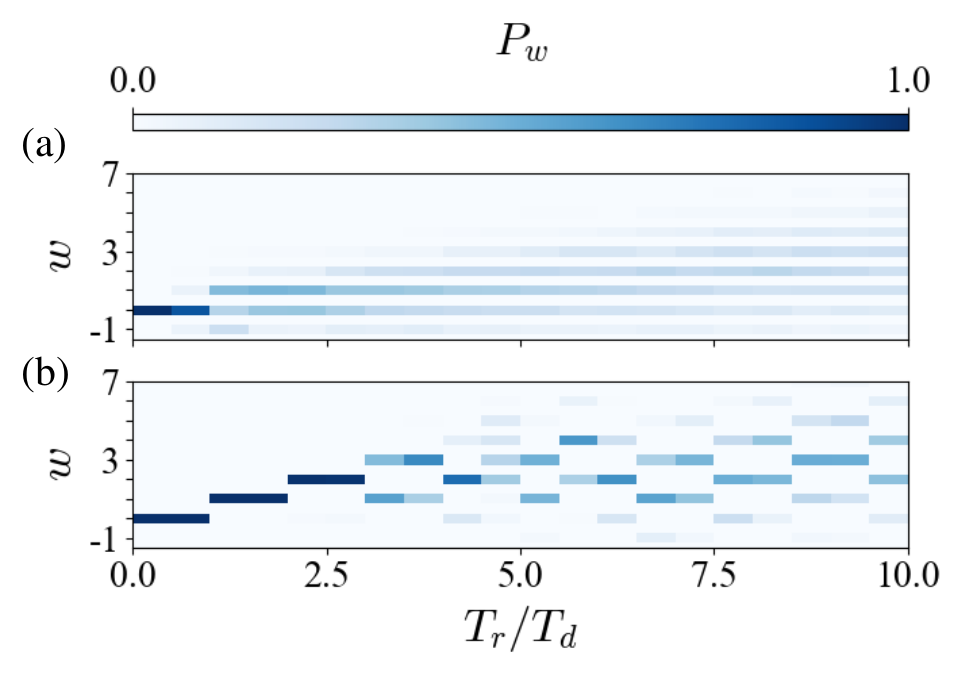}
\caption{(a), (b) Half-winding number distribution of the ODM under the switch-off protocol  for the driving parameters (a) $\{\lambda_{0}, \omega_{d}, A \} = \{ 0.9\lambda_{c}, 2\omega_{r, \mathrm{ODM}}, A_{r, \mathrm{ODM}} + 0.1 \}$ and (b) $\{\lambda_{0}, \omega_{d}, A \} = \{ 1.1\lambda_{c}, 0.8\omega, 0.55 \}$. The cavity dissipation for both figures is $\kappa = 1.0\omega$. }
\label{fig:switch_off}
\end{figure}

We can further explore the robustness of the bit-flip operation via quenches towards the SP by considering the case of $\omega_{d}' \rightarrow 0$, where the system is pushed back to the corresponding equilibrium phase prior to the driving. We present in Figs.~\ref{fig:switch_off}(a) and \ref{fig:switch_off}(b) the $P_{w}$ when $\omega_{d}'=0$ for $\lambda_{0} = 0.9 \lambda_{c}$, and $\lambda_{0} = 1.1 \lambda_{c}$, respectively. We demonstrate that for quenches back to the NP, $\lambda_{0}< \lambda_{c}$, the $P_{w}$ spreads out as $T_{r}$ increases, consistent with our results for quenches into the light-induced NP. On the other hand, for $\lambda_{0} > \lambda_{c}$, the $P_{w}$ finally localizes to either an even or odd integer of $w$, signaling that robust bit-flip operations can be performed by a sudden quench into the SP. In this case, the parameters considered lead to a steady state with macroscopic excitation of the cavity mode enough to suppress the quantum fluctuations.

Our results here demonstrate that bit-flip operations can be implemented by quenching the system from a DTC to another phase and then bringing the system back to a DTC after some time has elapsed. This bit-flip operation becomes even more robust when the system is quenched into a phase that has enough macroscopic excitations to counteract the effects of the quantum noise. Note that this mechanism also explains the robustness of the bit-flip operations for both the period-doubled and DTC states using the protocol in Eq.~\eqref{eq:phase_ramp_protocol} for $T_{\delta} \gg T_{d}$.

\section{Summary and Discussion}
\label{sec:conclusion}

In this work, we have proposed that a protocol based on introducing a defect in the phase of the drive \cite{apffel_experimental_2024} can be utilized for bit-flip operations of classical bits encoded in the period-doubled and DTC states of open systems. Furthermore, we have demonstrated the robustness of the bit-flip protocol against thermal and quantum noise.

We primarily focused on two systems: (i) a classical parametric oscillator connected to a thermal bath and; (ii) a quantum system of two-level systems or qubits coupled to a single photonic mode described by the open Dicke model. The ODM allows for investigating two different phases dependent on the strength of the light-matter interaction. For $\lambda_{0} < \lambda_{c}$, the steady state of the ODM is the NP, which can be approximated as a coupled PO when periodically driven. Meanwhile, for $\lambda_{0} > \lambda_{c}$, the ODM spontaneously breaks its $\mathbb{Z}_{2}$ symmetry in the static limit, leading to the breakdown of its coupled POs picture. The latter case allows us to explore bit-flip operations beyond the framework of parametric oscillators.

We have demonstrated that in the noiseless limit, which corresponds to the zero-temperature limit ($T \rightarrow 0$) for the PO and the thermodynamic limit ($N \rightarrow \infty$) for the ODM with $\lambda_{0} < \lambda_{c}$, the bit-flip operations become robust with increasing dissipation strength. In particular, for weak dissipation, their respective bit-flip diagrams exhibit fractal structures that can be attributed to the two systems entering transient irregular dynamics after applying the defect protocol. These fractal structures vanish for strong dissipation, allowing for robust switching in a large area of the parameter space spanned by $T_{\delta}$ and $A$. In the presence of fluctuations, the bit-flip operation remains robust against fluctuations when $T_{\delta}$ is sufficiently far from the critical defect duration needed for a successful bit-flip. Also, for $T_{\delta} \lessapprox T_{\delta, c}$, the half-winding number reveals that the defect protocol can have a non-zero probability to switch the period-doubled and DTC states, leading to a crossover-like behavior of the switching probability near $T_{\delta, c}$. This is in stark contrast with the discontinuous transition observed in noiseless POs \cite{apffel_experimental_2024}.

For the case of $\lambda_{0} > \lambda_{c}$ in the ODM, we have shown that the system can be pushed into various phases when the defect is switched on for a sufficient time. This has motivated us to explore the possibility of implementing bit-flip operations by quenching the driving frequency with variable defect duration $T_{r}$. We have shown that robust bit-flip operations are still possible if we choose an appropriate $\omega_{d}'$ that quenches the system into a new phase with a macroscopic occupation number enough to suppress the effects of noise from the dissipative channel. This mechanism is also responsible for protecting the bit-flip operation for large $T_{\delta}$ for both the thermal PO and the ODM for $\lambda_{0} < \lambda_{c}$.

Our work provides a general framework for implementing bit-flip operations on period-doubled and DTC states on systems where thermal and quantum fluctuations become relevant to the system's dynamics. We have extended the notion of defect protocols based on ramping the phase of the drive and have addressed subtle points concerning the specific form of the protocol for robust bit-flips. Concomitantly, we have provided a general method to dynamically manipulate DTCs, which can be of interest for more in-depth exploration of their nature, and can be used as a standard tool for potential applications of DTCs such as, but not limited to, quantum metrology \cite{moon_discrete_2024, yousefjani_discrete_2025, iemini_floquet_2024} and quantum computing \cite{bao_creating_2024}. Our results can be tested in existing platforms emulating the parametric oscillators \cite{leuch_parametric_2016, frimmer_rapid_2019, nosan_gate_controlled_2019}, and also systems that can emulate the open Dicke model, such as the atom-cavity system \cite{kesler_observation_2021, baumann_dicke_2010, klinder_dynamical_2015}, ensembles of nitrogen-vacancy centers \cite{amsuss_cavity_2011, astner_coherent_2017}, and cavity-magnon systems \cite{kim_observation_2025}, to name a few.

A natural extension of our work is to explore the effects of finite temperature in the robustness of the bit-flip operation on the DTC states. Also, given the generality of the protocol, it will be interesting to test this in systems that can emulate higher-order DTCs, in which the period of the DTC is greater than twice of the driving period, resulting to multiple degenerate DTC states. Higher-order DTCs have been explored in bosonic systems \cite{pizzi_period_2019, giergiel_creating_2020} and spin systems such as the Lipkin-Meshkov-Glick model \cite{pizzi_higher-order_2021}, and its generalization the $p$-spin model \cite{munoz-arias_floquet_2022, giachetti_fractal_2023}. This dynamical phase has also been observed experimentally in Rydberg atoms \cite{liu_higher-order_2024} and Kerr resonators \cite{taheri_all-optical_2022}. It will also be interesting to know the physical interpretation behind the half-winding number associated with the bit-flips, and whether one can relate it to topological phenomena that may arise in higher-order DTCs, in spirit of the topological interpretation of $w$ in Refs.~\cite{apffel_experimental_2024, simula_topological_2024}.

\section*{Acknowledgment}
This work was funded by the DOST-SEI Accelerated Science and Technology Human Resource Development Program.

\appendix

\section{Equations of motion of the ODM}\label{sec:eom}

We obtain the equations of motion of the ODM associated with the Heisenberg-Langevin equation in Eq.~\eqref{eq:heisenberg_langevin} by first evaluating the commutators between the Hamiltonian $\hat{H}$ and the cavity and collective spin operators, and then replacing the cavity mode operator with a complex number, $\hat{a} \rightarrow a \in \mathbb{C}$, and the collective spin operators with a real number, $\hat{S}^{x, y, z} \rightarrow S^{x, y, z} \in \mathbb{R}$. This leads to a set of stochastic differential equations of the form,
\begin{subequations}
\label{eq:open_dm_eom}
    \begin{equation}
        \partial_{t} a = -i \left( \omega a + \frac{2\lambda}{\sqrt{N}} S^{x} \right) - \kappa a + \xi(t), 
    \end{equation}
	\begin{equation}
		\partial_{t} S^{x} = - \omega_{0}S^{y},
	\end{equation}	    
    \begin{equation}
    		\partial_{t} S^{y} = \omega_{0} S^{x} - \frac{2\lambda}{\sqrt{N}} \left( a^{*} + a \right) S^{z},    
    \end{equation}
    \begin{equation}
  \partial_{t} S^{z} = \frac{2\lambda}{\sqrt{N}}\left( a^{*} + a \right) S^{y}.
    \end{equation}
\end{subequations}
To establish the thermodynamic limit, we substitute the rescaled quantities $\alpha = a / \sqrt{N}$ and $s^{x, y, z} = S^{x, y, z} / N$ back to Eq.~\eqref{eq:open_dm_eom}, which yields a rescaled equations of motion for the cavity and collective spins,
\begin{subequations}
\label{eq:rescaled_open_dm_eom}
    \begin{equation}
        \partial_{t} \alpha = -i \left( \omega \alpha + 2\lambda s^{x} \right) - \kappa \alpha + \frac{\xi(t)}{\sqrt{N}}, 
    \end{equation}
	\begin{equation}
		\partial_{t} s^{x} = - \omega_{0}s^{y},
	\end{equation}	    
    \begin{equation}
    		\partial_{t} s^{y} = \omega_{0} s^{x} - 2\lambda \left( \alpha^{*} + \alpha \right) s^{z},    
    \end{equation}
    \begin{equation}
	 \partial_{t} s^{z} = 2\lambda\left( \alpha^{*} + \alpha \right) s^{y}.
    \end{equation}
\end{subequations}
In this form, we can readily observe that the noise term has been renormalized such that $\xi(t) \rightarrow \tilde{\xi}(t) = \xi(t) / \sqrt{N}$. As such, if we take the thermodynamic limit, $N \rightarrow \infty$, the quantum fluctuation becomes negligible, allowing us to obtain the system's dynamics using mean-field methods.

For finite $N$, however, the contributions of the quantum noise in the dynamics become important. To capture the quantum noise, we employ semiclassical approximations based on phase-space methods, namely TWA and DTWA. For the TWA, instead of Eq.~\eqref{eq:open_dm_eom}, we numerically integrate the stochastic differential equation
\begin{subequations}
	\label{eq:twa_eom}
	\begin{equation}
	da_{\mathrm{R}} = \left(\omega a_{\mathrm{I}} - \kappa a_{\mathrm{R}} \right)dt  + \sqrt{\frac{\kappa}{2}}dW_{1}
	\end{equation}
	\begin{equation}
	da_{\mathrm{I}} = -\left( \omega a_{\mathrm{R}} + \frac{2\lambda}{\sqrt{N}}S^{x} + \kappa a_{\mathrm{I}} \right) + \sqrt{\frac{\kappa}{2}}dW_{2}
	\end{equation}
	\begin{equation}
		\partial_{t} S^{x} = - \omega_{0}S^{y},
	\end{equation}	    
    \begin{equation}
    		\partial_{t} S^{y} = \omega_{0} S^{x} - \frac{4\lambda}{\sqrt{N}} a_{\mathrm{R}} S^{z},    
    \end{equation}
    \begin{equation}
  \partial_{t} S^{z} = \frac{4\lambda}{\sqrt{N}}a_{\mathrm{R}} S^{y},
	\end{equation}
\end{subequations}
where we expand the cavity mode in terms of its real and imaginary component, $a = a_{\mathrm{R}} + ia_{\mathrm{I}}$. In this form, the quantum fluctuations are captured by the two independent Wiener processes $W_{1}$ and $W_{2}$. They both satisfy the conditions $\left<dW_{i} \right> = 0$ and $\left< dW_{i}dW_{j} \right> = \delta_{i, j}dt$ for $i = 1, 2$. For the DTWA, we instead consider the stochastic differential equations associated with each of the individual spins, $\vec{\sigma}_{i}$, appearing in Eq.~\eqref{eq:twa_eom} by writing $S^{x, y, z} = \sum_{i}^{N} \sigma_{i}^{x, y, z} / 2$. This leads to the following equations of motion for the DTWA:
\begin{subequations}
	\label{eq:dtwa_eom}
	\begin{equation}
	da_{\mathrm{R}} = \left(\omega a_{\mathrm{I}} - \kappa a_{\mathrm{R}} \right)dt  + \sqrt{\frac{\kappa}{2}}dW_{1}
	\end{equation}
	\begin{equation}
	da_{\mathrm{I}} = -\left( \omega a_{\mathrm{R}} + \frac{\lambda}{\sqrt{N}}\sum_{i}^{N} \sigma_{i}^{x} + \kappa a_{\mathrm{I}} \right) + \sqrt{\frac{\kappa}{2}}dW_{2}
	\end{equation}
	\begin{equation}
		\partial_{t} \sigma^{x}_{i} = - \omega_{0}\sigma^{y}_{i},
	\end{equation}	    
    \begin{equation}
    		\partial_{t} \sigma^{y}_{i} = \omega_{0} \sigma^{x}_{i} - \frac{4\lambda}{\sqrt{N}} a_{\mathrm{R}} \sigma^{z}_{i},    
    \end{equation}
    \begin{equation}
  \partial_{t} \sigma^{z}_{i} = \frac{4\lambda}{\sqrt{N}}a_{\mathrm{R}} \sigma^{y}_{i}.
	\end{equation}
\end{subequations}

\section{Regimes with stable period-doubled and DTC states}\label{sec:regime_of_stable_dtc}

\subsection{Resonance conditions for the parametric oscillator}\label{subsec:resonance_condition}

\begin{figure}
\centering
\includegraphics[scale=0.45]{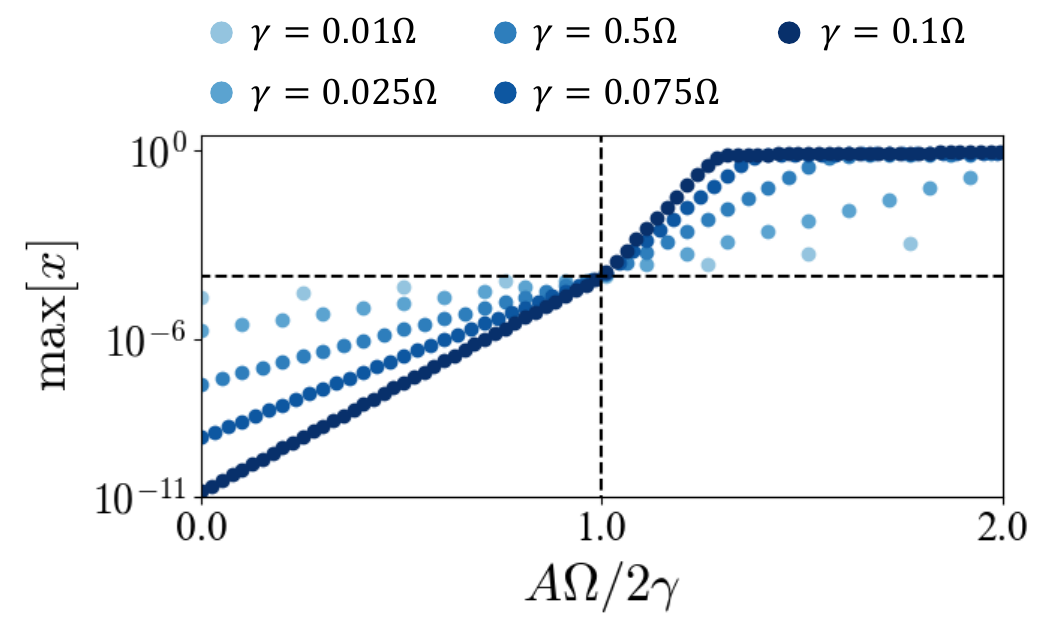}
\caption{Maximum value of $x$ within the interval $t \in \left[100T_{d}, 200 T_{d} \right]$ for $\omega_{d} = 2\Omega$ and different values of $\gamma$. The vertical dashed line marks the critical $A_{r}$ for visually guide, while the horizontal dashed line marks $\max[x] = x_{0}$.}
\label{fig:resonance_condition}
\end{figure}

To identify the regimes in the PO with period-doubled states for $\omega_{d} = 2\Omega$, we obtain the maximum value of $x=\sin(u)$ in the long-time limit. We do this by first initializing the system according to Eq.~\eqref{eq:dpp_initial_state}. We then allow the PO to relax into a steady oscillating state within the time interval $t \in \left[0, 200 T_{d}\right]$. Finally, we obtain $\max[x]$ during the time interval $t \in \left[ 100 T_{d}, 200 T_{d}\right]$. If $\max[x] \geq x_{0} \approx 10^{-4}$, we consider the system to be in a period-doubled state, otherwise, the PO only has a decaying oscillation towards $x = 0$.

We present in Fig.~\ref{fig:resonance_condition} the $\max[x]$ of the PO as a function of the driving amplitude $A$ for different dissipation strength $\gamma$. We can observe that $\max[x]$ has two different scaling behaviors depending on whether it is on a parametric resonance or not. In particular, $\max[x]$ has a larger scaling exponent when it is in a period-doubled state, which is most apparent for $\gamma = 0.1\Omega$. The critical point separating these two regimes collapses when we consider the particular form of scaling for the driving amplitude
\begin{equation}
 A_{\mathrm{scaled}} = A\Omega/2\gamma.
\end{equation}
With this scaling, the critical point lines up at $A_{\mathrm{scaled}} = 1$. This suggests that when the PO is resonantly driven at $\omega_{d} = 2\Omega$, the minimum driving amplitude needed to push the system into a period-doubled state is given to be
\begin{equation}
A_{r} = 2\gamma /\Omega,
\end{equation}
which is consistent with the critical point obtained for POs with Kerr nonlinearity using time-averaging methods \cite{heugel_ising_2022, ameye_parametric_2025}.

\subsection{Stability of period-doubled and DTC states in the presence of noise}\label{subsec:stable_dtc_under_noise}

\begin{figure}
\centering
\includegraphics[scale=0.333]{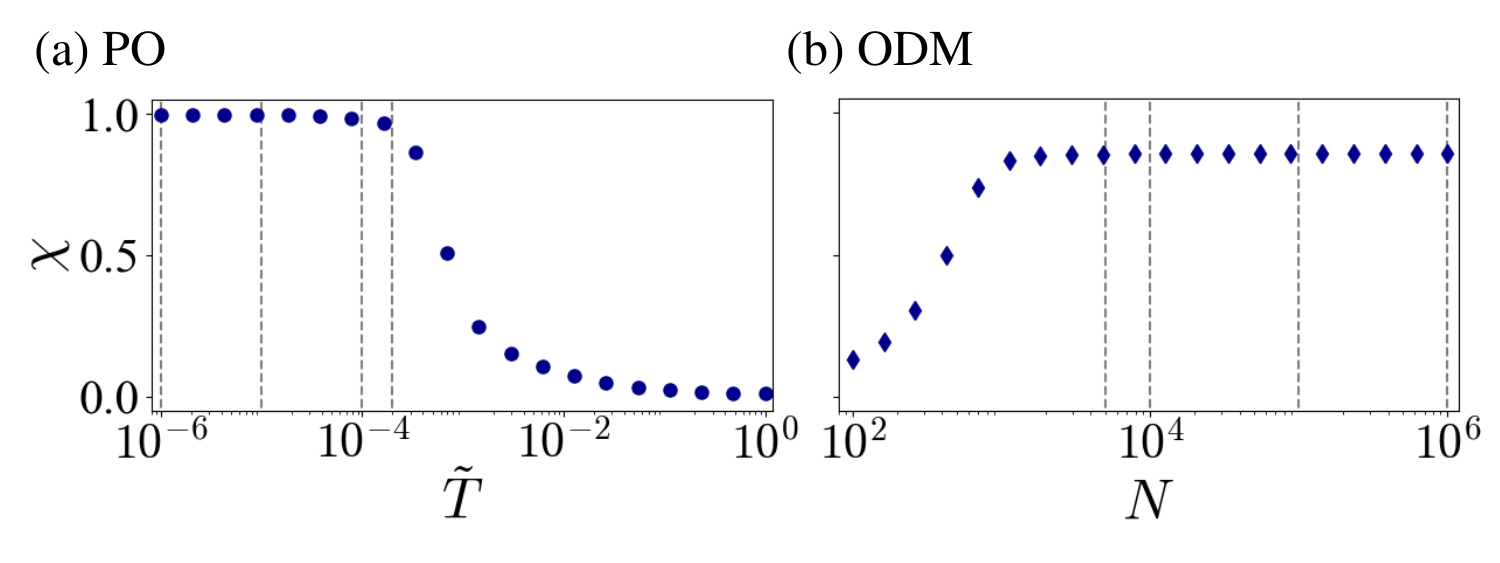}
\caption{(a), (b) Crystalline fraction of (a) the PO and (b) the ODM for $\lambda_{0} < \lambda_{c}$ as a function of $\tilde{T}$ and $N$ respectively. The parameters considered for PO are $\left\{ \gamma, \omega_{d}, \delta A \right\} = \left\{0.1\Omega, 2\Omega, 0.4\right\}$, while we consider $\left\{\kappa, \omega_{d}, \delta A \right\} = \left\{1.0\omega, 2\omega_{-}, 0.1 \right\}$ for the ODM. The vertical dashed lines mark the values of $\tilde{T}$ and $N$ considered in the main text.}
\label{fig:crystalline_fraction}
\end{figure}

Noise can diminish the coherence of the period-doubled and DTC states depending on its strength. This can either lead to thermally or quantum-activated switching between states \cite{marthaler_switching_2006} or the oscillations being washed out by strong fluctuations. To identify the regimes in which we can still test the robustness of bit-flip operations against noise without the need to account for these adverse effects, we quantify the quality of the oscillations associated with the period-doubled and DTC states using the crystalline fraction $\chi$ defined as
\begin{equation}
	\chi = \frac{P(\omega_{d} / 2)}{\int_{-\infty}^{\infty} P(\omega') d\omega'},
\end{equation} 
where $P(\omega)$ is the power spectrum of the order parameter considered. Here, the relevant order parameters are the $x$-quadrature for the thermal PO, and $S^{x}$ for the ODM.

We present in Fig.~\ref{fig:crystalline_fraction} the crystalline fraction of the thermal PO and ODM as a function of $\tilde{T}$ and $N$, respectively. We find that for small $\tilde{T}$ and large $N$, the thermal PO and the ODM have a crystalline fraction of $\chi \approx 1$, indicating that the period-doubled and DTC states at those regimes do not switch randomly and uncontrollably to their symmetry broken pairs due to strong noise. For $N = 1/\tilde{T} \leq 10^{3}$, the crystalline fraction decreases, which means that the bit-flip operations cannot be tested in these values of $N$ and $\tilde{T}$ as they coincide with the noise-activated switching of the period-doubled and DTC states. As such, in the main text, we only considered noise strengths marked by the gray dashed lines, all of which are in the regime where $\chi \approx 1$. Note that in the case of the DTC states, the crystalline fraction does not approach $\chi=1$ in the thermodynamic limit due to the presence of higher harmonics that are integer multiples of its dominant frequency, $\omega_{d} / 2$ \cite{jara_jr_theory_2024}.

\section{Fractal structures in bit-flip diagrams for weak dissipation}\label{sec:fractals}

\begin{figure}
\centering
\includegraphics[scale=0.48]{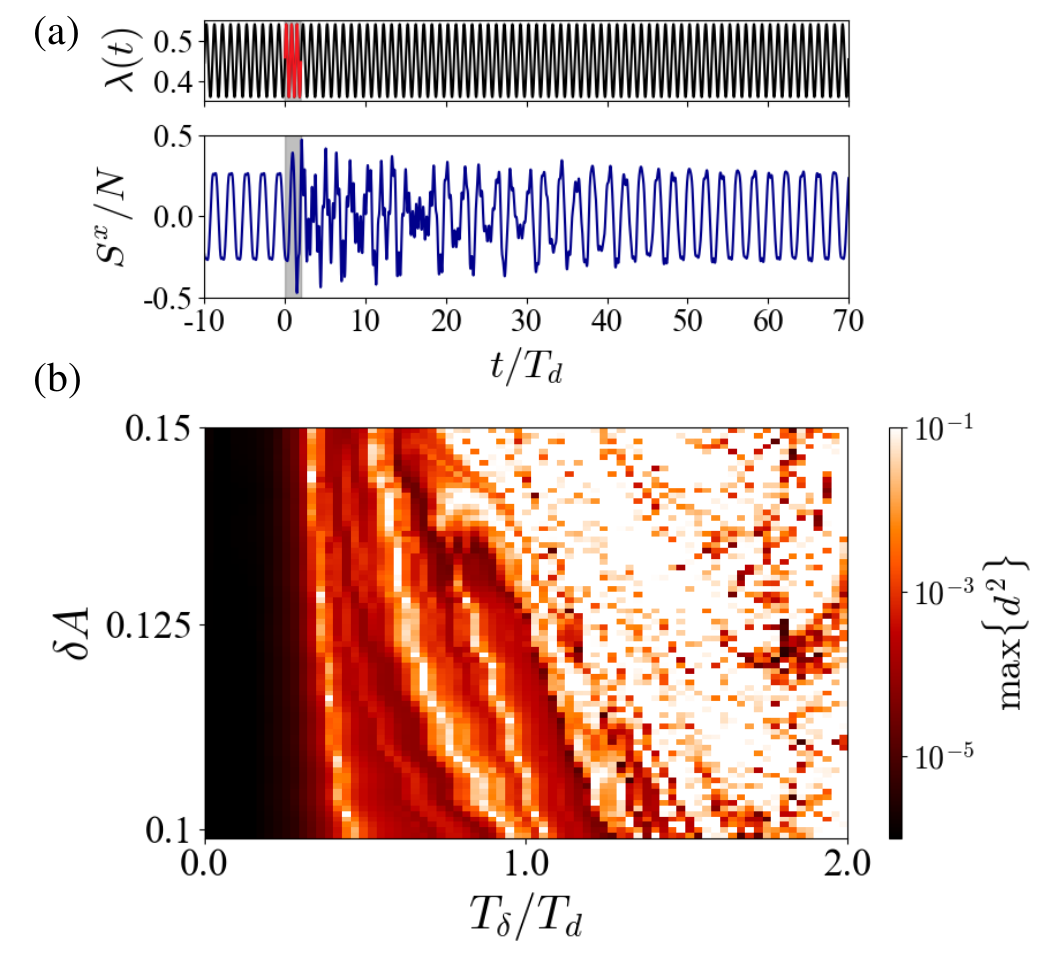}
\caption{(a) (Top panel) Defect protocol for $T_{\delta} = 2.0 T_{d}$. (Bottom panel) Exemplary dynamics of $S^{x} / N$ for the parameters $\kappa = 0.01\omega$, $\omega_{d} = 0.6\omega$, $A = 0.2$, and $T_{\delta} = 2.0 T_{d}$. The gray region corresponds to the time window at which the defect protocol is on. (b) Spin decorrelator of the ODM as a function of $T_{\delta}$ and $\delta A$ for the same parameters as Fig.~\ref{fig:switch_diagram}(g). }
\label{fig:decorrelator}
\end{figure}

To understand the emergence of fractal regions in the bit-flip diagrams and the fragility of the bit-flip operation for weak dissipation, we consider the dynamics of the ODM during the defect protocol. We show an exemplary dynamics of this scenario in Fig.~\ref{fig:decorrelator}(a), where we plot $S^{x}$ for $\kappa = 0.01 \omega$ and $T_{\delta} = 2.0 T_{d}$. Notice that when we apply the defect protocol, the system does not immediately return to the DTC phase after $t = T_{\delta}$. Instead, due to the weak dissipation, the system exhibits transient irregular dynamics that persists for sufficiently long time until it relaxes back to a DTC. This irregularity of the dynamics leads to strong dependence of the final state of the DTC on the system's fractal basin of attraction \cite{Tel_Gruiz_2006}. We confirm this by quantifying the degree of irregularity in the dynamics of the ODM for a given $\delta A$ and $T_{\delta}$ using the decorrelator \cite{pizzi_higher-order_2021}
\begin{equation}
\label{eq:spin_decorrelator}
d^{2}(t) = ||\vec{s}_{\mathrm{O}} - \vec{s}_{\mathrm{P}} ||^{2} = \frac{1}{2} - 2  \left( \vec{s}_{\mathrm{O}} \cdot \vec{s}_{\mathrm{P}} \right),
\end{equation}
where $\vec{s}_{\mathrm{O}, \mathrm{P}} = \begin{pmatrix}
S^{x}_{\mathrm{O}, \mathrm{P}} & S^{y}_{\mathrm{O}, \mathrm{P}} & S^{z}_{\mathrm{O, P}} 
\end{pmatrix}^{T} / N
$, and by the conservation of spin angular momentum for the ODM \cite{emary_chaos_2003}, $||\vec{s}_{O}||^{2} = ||\vec{s}_{P}||^{2} = \frac{1}{4}$. The decorrelator determines the deviation between two identical systems, labeled as systems $\mathrm{O}$ and $\mathrm{P}$, initialized in a nearly identical state \cite{pizzi_higher-order_2021}. In particular, for regular dynamics, we expect that $\vec{s}_{\mathrm{O}}$ and $\vec{s}_{\mathrm{P}}$ will be aligned at all times and thus $d^{2}(t) = 0$. Meanwhile, we get the maximum $d^{2} = 1$ when $\vec{s}_{\mathrm{O}}$ and $\vec{s}_{\mathrm{P}}$ are anti-aligned with one another.

We calculate the $d^{2}$ associated with the transient irregular dynamics after the defect as follows. We consider two identical systems that are represented by their respective collective spin vectors $\vec{s}_{\mathrm{O}}$ and $\vec{s}_{\mathrm{P}}$, and cavity modes, $a_{\mathrm{O}}$ and $a_{\mathrm{P}}$. At $t_{i} = -100 T_{d}$, we prepare the two systems in the initial state given by Eq.~\eqref{eq:dm_initial_state} and allow it to evolve in an identical DTC state. We then perturb the system $\mathrm{P}$ at $t=0$ such that $a_{\mathrm{P}}(t = 0) = a_{\mathrm{O}}(t = 0) + \epsilon_{a}$, where $\epsilon_{a} = 1$. To also observe the sensitivity of the dynamics with the driving amplitude, we quench the driving amplitude of the system $\mathrm{P}$ into $A_{\mathrm{P}} = A_{\mathrm{O}} + \epsilon_{A}$, with $\epsilon_{A} = 10^{-5}$. We finally apply defect protocol on the two systems within the time interval $t \in \left[0, T_{\delta} \right]$ and observe how the deviation between $\vec{s}_{\mathrm{O}}$ and $\vec{s}_{\mathrm{P}}$ grows in time. We identify the irregular dynamics as solutions having $\max\{ d^{2}(t) \} \geq 10^{-1}$. We present in Fig.~\ref{fig:decorrelator}(b) the $\max\{d^{2}\}$ as a function of $\delta A$ and $T_{\delta}$ for the same set of parameters as Fig.~\ref{fig:switch_diagram}(g). We find that $\max\{d^{2} \}$ is small in regions where we do not observe fractal-like boundaries. This is in contrast to regions with large $T_{\delta}$, wherein the fractal structures coincide with areas where $\max\{ d^{2} \} \geq 10^{-1}$. Therefore, we confirm that the fractal structures in the bit-flip diagrams of the two systems can be attributed to the transient irregular dynamics for weak dissipation.

\section{Defect protocol with phase error}\label{sec:defect_with_phase_error}

\begin{figure}
\centering
\includegraphics[scale=0.45]{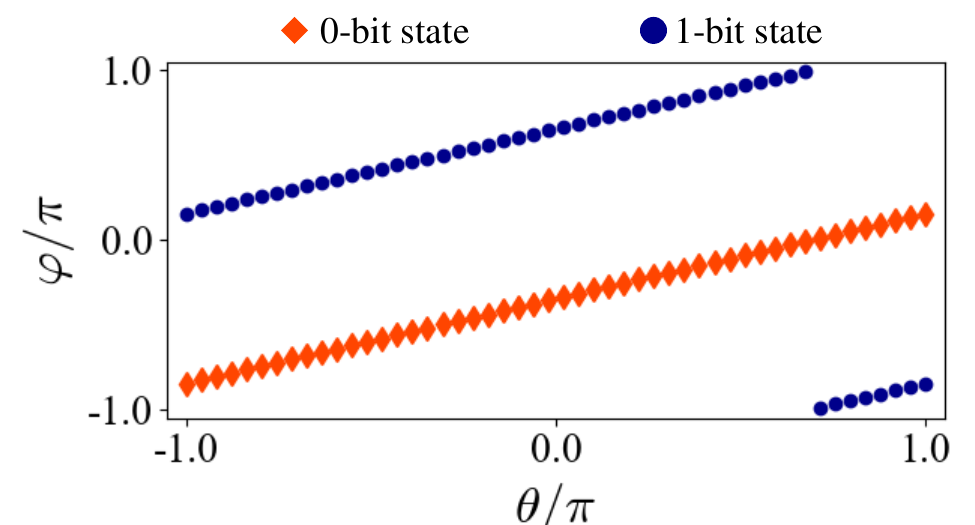}
\caption{Absolute-time phase corresponding to the two degenerate DTC states of the ODM as a function of the phase of the drive $\theta$. The driving parameter considered is $\left\{ \kappa, \omega_{d}, A \right\} = \left\{1.0\omega, 0.8\omega, 0.55 \right\}$. Note that the labeling of the DTC states is arbitrary, hence, for this particular example, we label the state with the smaller $\varphi$ as the $0$-bit state.}
\label{fig:dtc_states}
\end{figure}

All of the defect protocols considered in the main text involve the phase of the drive going back to zero after the defect. We address here what happens when the driving protocol does not go back to its original state after the defect protocol and instead obtains a phase shift $\theta_{f}$. We begin by determining the relationship between the degenerate DTC states and the phase of the driving protocol, $\theta$, in the absence of any defect. As shown in Fig.~\ref{fig:dtc_states}, the absolute-time phase $\varphi$ of the two degenerate DTC states follows a linear trend with $\theta$, indicating that $\theta$ dictates the possible values of $\varphi$. This can be understood by viewing the driving protocol as a periodic potential, in which the oscillations of the DTC are pinned. As such, the $\varphi$ of the two degenerate DTC states will follow the phase of the driving protocol accordingly.

\begin{figure}
\centering
\includegraphics[scale=0.60]{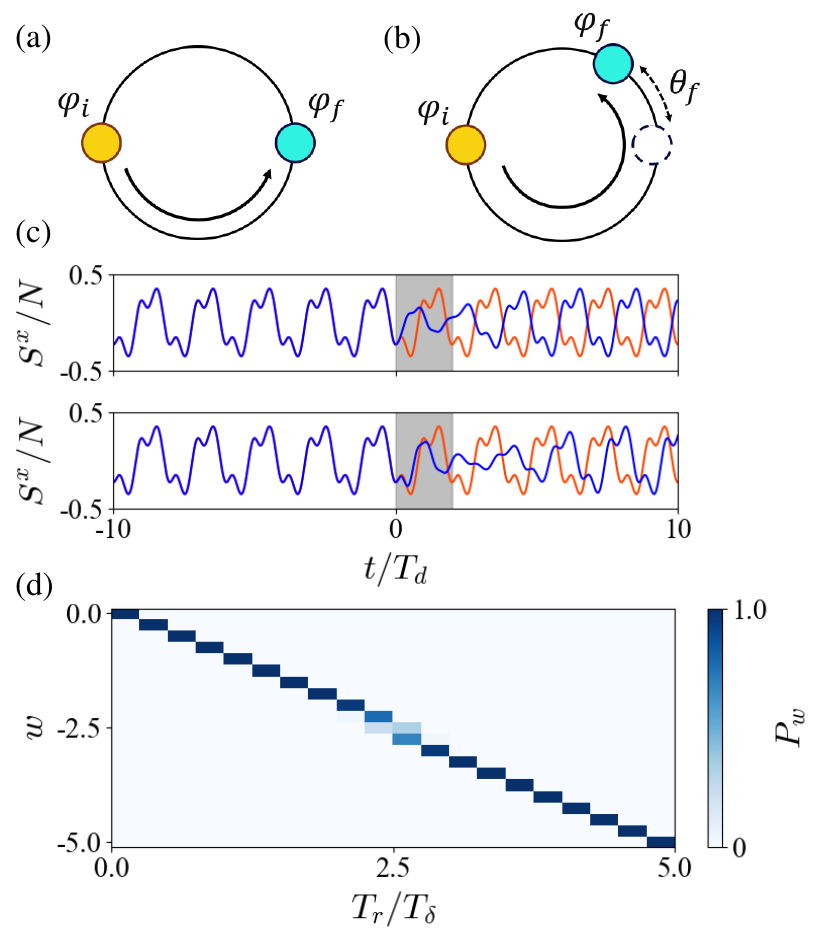}
\caption{(a), (b) Sketch of the bit-flip operation when the driving protocol has a phase error of (a) $\theta_{f} = 0$, and (b) $\theta_{f} \neq 0$ after the defect protocol. (c) (Top panel) Exemplary period-doubled state switching when the driving protocol has a phase error $\theta_{d} = \pi/ 2$ only during the defect protocol. (Bottom panel) Bit-flip when the driving protocol has a phase error of $\theta_{f} = \pi/2 $ after the defect protocol. For both dynamics, the parameters are $\omega_{d} = 0.8$, $A = 0.55$, $T_{r} = T_{\delta} = 2.0 T_{d} $.(d) Half-winding number distribution for the ODM under the continuous quenched-frequency protocol for the same parameters as Fig.~\ref{fig:quenched_frequency}(b).}
\label{fig:phase_error}
\end{figure}

The dependence between $\varphi$ and the phase of the drive affects the way the DTC switches states when we consider a general defect protocol of the form,
\begin{equation}
\lambda(t) = 
\begin{cases}
\lambda_{0}\left[1 + A\sin(\omega_{d} t + \theta_{i}) \right], \quad \quad t \leq 0\\
\lambda_{0} \left[1 + A\sin(\omega_{d}' t + \theta_{D}) \right], \quad \quad 0 < t < T_{r} \\
\lambda_{0} \left[1 + A\sin(\omega_{d}t + \theta_{f}) \right], \quad \quad t \geq T_{r}. 
\end{cases}
\end{equation}
where $\theta_{i}$, $\theta_{d}$, and $\theta_{f}$ are the phases of the drive before, during, and after the defect protocol, respectively. Consider for instance $T_{r} = T_{\delta}$ such that $\omega_{d}'$ is only dictated by the defect duration. Without loss of generality, we will set $\theta_{i} = 0$. Suppose we consider a suitable $T_{\delta}$ such that the DTC switches after the defect when $\theta_{D} = \theta_{f} = 0$. Given that the $\varphi$ of the two degenerate DTC states depends on the phase of the drive, we can infer that at $t > T_{r}$, the system does not perfectly switch into one of the original DTC states when $\theta_{f}\neq 2\pi n$ with $n \in \mathbb{Z}$, as shown in Fig.~\ref{fig:phase_error}(b). Instead, the system incurs an absolute-time phase of $\varphi = \pi + \theta_{f}$, resulting in an imperfect bit-flip relative to the original bit encoded in the period-doubled or DTC state. This behavior also implies that $\theta_{d}$ does not affect the overall bit-flip dynamics of the system. We verify this in Fig.~\ref{fig:phase_error}(c), where we consider the cases of $\{ \theta_{D}, \theta_{f} \} = \{ \pi / 2, 0 \}$ and $\{ \theta_{D}, \theta_{f} \} = \{ 0, \pi / 2 \}$, respectively. As we can observe, we obtain perfect switching even when $\theta_{d} \neq 0$ provided $\theta_{f} = 0$. As soon as $\theta_{f}$ becomes nonzero, the DTC obtains a phase shift of $\varphi = \pi + \theta_{f}$, which leads to an imperfect switching of a period-doubled or DTC state.

Given that the bit-flip operation of DTC states only depends on $\theta_{f}$, we now present in Fig.~\ref{fig:phase_error}(d) the $P_{w}$ of the ODM for the case of $T_{r} \neq T_{\delta}$ and $\theta_{f} =2\pi T_{r} / T_{\delta} $, with $\theta_{D} = 0$. Unlike the driving protocol described in Eq.~\eqref{eq:quenched_frequency}, this choice of $\theta_{f}$ preserves the continuity of the drive throughout the defect protocol. We use the same parameters as in Fig.~\ref{fig:quenched_frequency}(b) to construct Fig.~\ref{fig:phase_error}(d). In contrast to the behavior of $P_{w}$ in Fig.~\ref{fig:quenched_frequency}(b), we find in Fig.~\ref{fig:phase_error}(d) that $P_{w}$ has a continuous dependence with $\theta_{f}$, with the most probable $w$ following the linear trend $w = -\theta_{f}/2\pi = -T_{r}/T_{\delta}$, suggesting the sensitivity of the bit-flip operation in the final phase of the drive. Note that these results should apply to the thermal PO as well since the same principles apply to them, i.e. the dependence of the degenerate period-doubled states on the driving phase.

\bibliography{reference.bib}

\end{document}